\documentclass[11pt]{article}
\usepackage{latexsym}
\usepackage{amsmath,amsfonts,amsbsy}
\usepackage[dvips]{graphicx,epsfig}

 \csname
@addtoreset\endcsname{equation}{section}

\def\mso{\mathfrak{so}}

\def\mhso{\mathfrak{ho}}

\def\Real{{\mathbb R}}
\def\Comp{{\mathbb C}}
\def\integ{{\mathbb Z}}
\def\id{{\mathbb I}}

\def\bec{\begin{center}}
\def\ec{\end{center}}
\def\a{\alpha} \def\ad{\dot{\a}} 

\def\b{\beta}  \def\bd{\dot{\b}} 
\def\c{\gamma} \def\cd{\dot{\c}}
\def\C{\Gamma}
\def\d{\delta} \def\dd{\dot{\d}}

\def\e{\epsilon}

\def\k{\kappa}
\def\l{\lambda}

\def\m{\mu}
\def\n{\nu}
\def\r{\rho}
\def\s{\sigma}

\def\t{\tau}
\def\th{\theta} 

\def\x{\xi}
\def\y{\eta}

\def\O{\Omega}
\def\o{\omega}

\def\sb{{\bar\s}}

\def\yb{{\bar y}}
\def\zb{{\bar z}}

\def\ob{{\bar \o}}

\def\ts{\tilde{\s}}

\let\la=\label

\def\nn{\nonumber}
\newcommand{\eq}[1]{(\ref{#1})}

\newcommand{\w}[1]{\\[0.#1cm]}
\def\be{\begin{equation}}
\def\ee{\end{equation}}
\def\bea{\begin{eqnarray}}
\def\eea{\end{eqnarray}}
\def\ba{\begin{array}}
\def\ea{\end{array}}

\def\mx#1#2#3#4{\left#1\begin{array}{#2} #3 \end{array}\right#4}

\def\ft#1#2{{\textstyle{{\scriptstyle #1}
\over {\scriptstyle #2}}}}

\def\scs#1{\section{\sc \large #1}}
\def\scss#1{\subsection{\sc  #1}}
\def\scsss#1{\subsubsection{\sc \small #1}}

\def\ad{\dot\alpha}
\def\bd{\dot\beta}
\def\sb{\bar\sigma}

\begin{document}
\begin{flushright}
ROM2F/2007/10\\
MIFP-07-16\\
hep-th/yymmddd \vskip 8pt {\today}\\

\end{flushright}

\vspace{8pt}

\begin{center}


{\Large\sc Real Forms of Complex Higher Spin
\\[10pt]Field Equations and New Exact Solutions}


\vspace{12pt}

Carlo~Iazeolla${}^{1,2}$, Ergin~Sezgin${}^{3}$ and Per~Sundell${}^{2}$\\[15pt]

{\small${}^{1}${\it{Dipartimento di Fisica, Universit\`{a} di Roma
``Tor
Vergata"}}\\
     {\it{INFN, Sezione di Roma ``Tor Vergata" }}\\
      {\it{Via della Ricerca Scientifica 1, 00133 Roma, Italy }}\\[15pt]
${}^{2}${\it Scuola Normale Superiore and INFN\\
Piazza dei Cavalieri 7, 56100 Pisa, Italy}\\[15pt]
${}^3${\it George P. and Cynthia W. Mitchell Institute for
Fundamental Physics\\ Texas A\&M University\\College Station, TX
77843-4242, USA}}


\vspace{10pt} {\sc\large Abstract}\end{center}

We formulate four dimensional higher spin gauge theories in
spacetimes with signature $(4-p,p)$ and nonvanishing cosmological
constant. Among them are chiral models in Euclidean $(4,0)$ and
Kleinian $(2,2)$ signature involving half-flat gauge fields. Apart
from the maximally symmetric solutions, including de Sitter
spacetime, we find: (a) $SO(4-p,p)$ invariant deformations,
depending on a continuous and infinitely many discrete parameters,
including a degenerate metric of rank one; (b) non-maximally
symmetric solutions with vanishing Weyl tensors and higher spin
gauge fields, that differ from the maximally symmetric solutions in
the auxiliary field sector; and (c) solutions of the chiral models
furnishing higher spin generalizations of Type D gravitational
instantons, with an infinite tower of Weyl tensors proportional to
totally symmetric products of two principal spinors. These are
apparently the first exact 4D solutions with non-vanishing massless
higher spin fields.

\setcounter{page}{1}

\pagebreak

\tableofcontents

\newpage


\scs{Introduction}\label{sec:in}


Given the impact Yang-Mills theory and Gravity have had on the
development towards our present understanding of fundamental
interactions, as formulated within Quantum Field Theory (QFT) and to
some extent String Field Theory (SFT), it is natural to explore
higher spin (HS) extensions of gauge symmetries (\emph{i.e.}
non-abelian gauge groups containing generators in representations of
the Lorentz group with spins higher than one).

Presently, the only known full models of interacting higher-spin
gauge fields are those based on the Vasiliev equations
\cite{vasiliev}. These equations are naturally formulated in terms
of $SL(2;\Comp)$ spinor oscillators in Lorentzian signature $(3,1)$.
In this paper, we shall formulate them using spinor oscillators in
Euclidean signature $(4,0)$ and Kleinian signature $(2,2)$ as well,
and present nontrivial exact solutions with novel properties such as
the excitation of all higher spin fields. Before we state our
motivations for this work, let us first highlight some key elements
of the HS theory.

To begin with, the Vasiliev equations that describe the HS theory
involve two features that are relatively novel from the
point-of-view of lower-spin QFT as well as the standard formulation
of SFT. Firstly, they are written in a frame-like language, closely
related to the constraint formulation of supergravity, known as free
differential algebra (FDA), or unfolded dynamics. Here, \emph{all}
fields are differential forms, which live on an \emph{a priori}
unspecified base manifold. Moreover, for \emph{each} differential
form there is a, in general non-linear, differential constraint,
written using the exterior derivative (and no contractions of curved
indices using the metric). Thus, diffeomorphism invariance is
manifest without the need to single out a metric or other component
field.

Secondly, in order to accommodate an infinite number of physical as
well as auxiliary fields, one works with master fields that, in
addition to being differential forms, are functions of oscillator
variables. The functions belong to, or, depending on taste, define a
fiber over the base manifold consisting of representations of an
underlying non-abelian higher-spin algebra. In particular, the
master zero-form is directly related to the (massless) spectrum via
the theorem of Flato and Fronsdal \cite{Flato:1978qz}. One may go
further and associate the oscillators to a particle or other
extended objects, perhaps related to discretization of tensionless
strings and membranes in AdS \cite{Engquist:2007vj}, though these
considerations are of course not crucial for setting up Vasiliev's
formalism.

The simplest higher spin gauge theories of Vasiliev type, and indeed
the first ones to appear in the literature \cite{vasiliev}, are
based on higher-spin extensions of $SO(3,2)$ realized using
oscillators that are $SL(2,\Comp)$ doublets (coordinatizing the
phase space of Dirac's $Sp(4)$ singleton). Here, the {\it master
fields} are an adjoint {\it one-form} and a twisted-adjoint {\it
zero-form}, sometimes referred to as the Weyl zero-form. The master
field equations, which we again stress are manifestly background
independent and diffeomorphism invariant, can then be written on a
remarkably simple closed form. These equations can be treated in two
almost opposite ways, namely by projecting to the fiber or by
projecting to the base. In the latter case one can make contact with
lower-spin field theory by taking the base manifold to be an
ordinary spacetime and eliminate the auxiliary fields, treating
\emph{only} the Lorentz connection and the vierbein exactly. This
well-defined, albeit tedious, approach yields manifestly
reparametrization and locally Lorentz invariant physical field
equations in a perturbative expansion in curvatures as well as
higher-spin gauge fields.

The projection to the fiber, on the other hand, is a more tractable
operation, since the Vasiliev equations can be solved locally on the
base manifold using gauge functions \cite{Prokushkin:1998bq}. This
leaves equations on the non-commutative fiber, which are thus purely
algebraic from the point-of-view of the base manifold. The simplest
exact solution to these equations is the four-dimensional anti-de
Sitter spacetime. In a recent paper, \cite{Sezgin:2005pv}, we have
given an exact $SO(3,1)$-invariant solution to these equations. The
solution describes a locally time-dependent solution with a local
space-like singularity that can be resolved by the method of
patches. The solution is asymptotically AdS and periodic in time, so
that one may think of it as an ``instanton universe'' inside AdS
\cite{Sezgin:2005hf}.  More recently, the gauge function method has
been used to describe the BTZ black hole metric as a solution to
full three dimensional HS gauge theory \cite{Didenko:2006zd}.

This raises the question how to Wick rotate solutions of the
Lorentzian theory into solutions of a Euclidean theory. The main
difficulty is to impose proper reality conditions given the doubling
of the spinor oscillators due to the Euclidean signature. We resolve
this by taking the master fields to be holomorphic functions of the
left-handed and the right-handed spinor oscillators subject to
pseudo-reality conditions.

In addition to the Euclidean signature, we shall consider the
Kleinian signature as well. While in all signatures there is the
possibility of a chiral asymmetry, in Euclidean and Kleinian
signatures, the extreme case of parity violation involving half-flat
gauge fields can also arise. We refer to the latter ones as
\emph{chiral models}. In HS gauge theory, the HS algebra valued
gauge-field curvatures can be made, say, self-dual, but the model
nonetheless contains the anti-self-dual gauge fields through the
master zero-form which contains the corresponding Weyl tensor
obeying the appropriate field equation. Although this is contrary to
what happens in ordinary Euclidean gravity, where the field
equations can contain only self-dual fields, it is not a surprise in
higher spin theory since the underlying higher spin algebra, which
is an extension of $SO(5)$, does not admit a chiral massless
multiplet \footnote{We shall leave the group-theoretical analysis to
\cite{wip}, where we also give the spinor-oscillator formulations of
the four-dimensional minimal bosonic models with $H_4$, $dS_4$ and
$H_{3,2}$ vacua.}.

There are several reasons that make the investigation of HS theory
in Euclidean and Kleinian signatures worthwhile. To begin with, just
as the Euclidean version of gravity plays a significant role in the
path integral formulation of quantum gravity, it is reasonable to
expect that this may also be the case in the quantum formulation of
HS theory, despite the fact that an action formulation is yet to be
spelled out (see, \cite{Engquist:2007kz} for recent progress). For
reviews of Euclidean quantum gravity, see, for example, \cite{GH}
and \cite{Gibbons}.

Another well known  aspect of self-dual field theories is their
capability to unify a wide class of integrable systems in two and
three dimensions. It would be interesting to extend these
mathematical structures to self-dual HS gauge theories to find new
integrable systems.

The chiral HS theories in Kleinian signatures may also be of
considerable interest in closed $N=2$ string theory in which the
self-dual gravity in $(2,2)$ dimensions arises as the effective
target space theory \cite{Ooguri:1990ww}. However, there are some
subtleties in treating the picture-changing operators in the BRST
quantization which have raised the question of whether there are
more physical states \cite{Junemann:1997rx}, and in the case of open
$N=2$ theory an interpretation in terms of an infinite tower of
massless higher spin states has been proposed
\cite{Devchand:1997dq}. It would be very interesting to establish
whether these theories or their possible variants admit self-dual HS
theory in the target space. While the $N=2$ string theories may seem
to be highly unrealistic, it should not be ruled out that they may
be connected in subtle ways to all the other string theories which
themselves are connected by a web of dualities in M theory.

In this paper, we shall take the necessary first steps to start the
exploration of the Euclidean and Kleinian HS theories. We shall
start by determining the real forms of the higher spin algebra based
on infinite dimensional extension of $SO(5;\Comp)$ and formulate the
corresponding higher spin gauge theories in four-dimensional
spacetime with signature $(4-p,p)$. Maximally symmetric four-
dimensional constant curvature spacetimes, including de Sitter
spacetime, defined by the embedding into five-plane with signature
$(5-q,q)$ are readily exact solutions. Fluctuations about these
spaces arrange themselves into all the irreducible representations
of $SO(5-q,q)$ contained in the symmetric two-fold product of the
fundamental singleton representation of this group, each occurring
once. The details of this phenomenon will be provided in a separate
paper \cite{wip}.

We then devote the rest of the paper to finding a class of
nontrivial exact solutions of these models, including the Euclidean
and chiral cases. The key information about these solutions is
encoded in the master zero-form which contains a real ordinary
scalar field, and the {\it Weyl\ tensors} $\phi_{\a_1\cdots
\a_{2s}}$ and $\phi_{\ad_1,\cdots \ad_{2s}}$ for spin $s=2,4,6,...$
in the minimal bosonic model and $s=1,2,3,4,...$ in a non-minimal
bosonic model \cite{Vasiliev:1995dn,Engquist:2002vr}. Vasiliev's
full higher spin field equations assume a form reminiscent of that
of open string field theory, with master fields that are functions
of spacetime as well as an internal noncommutative space of
oscillators. Our new exact solutions are constructed by using the
oscillators to build suitable projectors, with slightly different
properties in the minimal and non-minimal models.

Our exact solutions fall into the following four classes:

\underline{Type 0}:

These are {\it maximally symmetric solutions} (see Table 1) with
\bea  \phi(x) &=& 0\ ,\qquad \Phi_{\a_1\cdots \a_{2s}}=0\ , \qquad
\Phi_{\ad_1\cdots\ad_{2s}}=0\ , \nn\w2
e_\mu^a &=& {4\delta_\mu^a\over (1-\lambda^2x^2)^2} \ ,
 \qquad W_\mu{}^{a_1\cdots a_{s-1}}=0\ , \label{type0}\eea
describing the symmetric spaces $S^4, H_4, AdS_4, dS_4,
H_{3,2}=SO(3,2)/SO(2,2)$, where $|\lambda|$ is the inverse radius of
the symmetric space, $x^2=x^a x^b\eta_{ab}$, and $\eta_{ab}$ is the
tangent space metric. In the above the zero-forms have spin
$s=2,4,6,...$ in the minimal model and $s=1,2,3,4,...$ in the
non-minimal model, while for $W_\mu{}^{a_1\cdots a_{s-1}}$,
$s=4,6,...$ in the minimal model, and $s=1,3,4,5,6,...$ in the
non-minimal model.

\underline{Type 1}:

These solutions, which arise in the minimal models (and therefore
are evidently solutions also to the non-minimal models with
vanishing odd spins), are {\it $SO(p,4-p)$ invariant deformations}
of the maximally symmetric solutions with
\bea  \phi(x) &=& \nu (1-\lambda^2 x^2)\ ,\qquad \Phi_{\a_1\cdots
\a_{2s}}=0\ , \qquad \Phi_{\ad_1\cdots\ad_{2s}}=0\ , \   \
(s=2,4,...) \nn\w2 e_\mu^a &=& f_1 \delta_\mu^a + \lambda^2 f_2
x_\mu x^a\ ,
 \qquad W_\mu{}^{a_1\cdots a_{s-1}}=0\ \ \ (s=4,6,...)\ , \eea
where $\nu$ is a continuous parameter and $f_1,f_2$ (see \eq{f12})
are highly complicated functions of $x^2$, $\nu$, and a set of
\emph{discrete} parameters corresponding to whether certain
projectors are switched on or off. The metric is Weyl-flat conformal
to the maximally symmetric solution with a complicated conformal
factor, and note that all the higher spin gauge fields vanish.
Interestingly, a particular choice of the discrete parameters yield,
in the $\nu \rightarrow 0$ limit, the {\it degenerate metric}:
\be g_{\mu\nu}= {1\over (1-\lambda^2 x^2)}{x_\mu x_\nu\over
\lambda^2 x^2}\ . \ee

Degenerate metrics are known to play a role topology change in
spacetime (see, for example, \cite{Horowitz:1990qb}, and references
therein). Interestingly, here they arise in a natural way by simply
taking a certain limit in the parameter space of our solution.

\underline{Type 2}:

These are solutions of the non-minimal model that are \emph{not}
solutions to the minimal model. The spacetime component fields are
identical to those of the maximally symmetric Type 0 solutions, but,
unlike in the Type 0 solution, the spinorial master one-form is
non-vanishing (see \eq{spinorform}). Even though all odd spin fields
are vanishing, the solution exists only for the non-minimal model
because the spinorial master field violates the kinematic conditions
of the minimal model. In particular, this means that this type of
solution cannot be a $\nu\rightarrow 0$ limit of the Type 1
solutions. Furthermore, the spinorial master field is parametrized
by {\it discrete} parameters, again associated with projectors.

\underline{Type 3}:

These are solutions of the {\it non-minimal chiral models} in
Euclidean and Kleinian signatures, in which {\it all gauge fields
are non-vanishing}. These solutions also depend on an infinite set
of {\it discrete parameters} and for simple choices of these
parameters we obtain  two such solutions in both of which
\be \phi(x) \ =\  -1\ , \qquad \Phi_{\a_1\cdots \a_{2s}}\ =\ 0 \
,\qquad W_\mu{}^{a_1\cdots a_{s-1}}\ \neq\ 0\ . \ee
In one of the solutions the Weyl tensors and the vierbein take the
form
\bea && \Phi_{\ad_1\cdots\ad_{2s}}\ = \ -2^{2s+1}(2s-1)!!\left(
{h^2-1 \over \e h^2}\right)^s\, U_{(\ad_1}\cdots U_{\ad_s}\,
V_{\ad_{s+1}}\cdots V_{\ad_{2s})}\ ,\w2
&& e_\mu^a\ =\ {-2\over h^2(1+2g)}\left[g_3\delta_\mu^a +g_4
\lambda^2 x_\mu x^a + g_5\lambda^2 (Jx)_\mu (Jx)^a\right]\ , \eea
where $h,g,g_3,g_4,g_5$ are functions of $x^2$ defined in \eq{aad},
\eq{g345}, and the almost complex structure $J_{ab}$ and spinors
$(U,V)$ are defined in \eq{jj} and \eq{UV}, and $\e=\pm 1$ as
explained in Section 3.5.
For the other solution we have
\bea && \Phi_{\ad_1\cdots\ad_{2s}}\ = \ -2^{2s+1}(2s-1)!!\left(
{1\over \e h^2}\right)^s\, \bar\l_{(\ad_1}\cdots \bar\l_{\ad_s}\,
\bar\mu_{\ad_{s+1}}\cdots \bar\mu_{\ad_{2s})}\ ,\w2
&& e_\mu^a\ =\ {-2\over h^2(1+2{\tilde g})}\left[ \delta_\mu^a
+{\tilde g}_4 \lambda^2 x_\mu x^a + {\tilde g}_5\lambda^2 ({\tilde
J}x)_\mu ({\tilde J}x)^a\right]\ , \eea
where the functions ${\tilde g}, {\tilde g}_4, {\tilde g}_5$ are
defined in \eq{tildeg345}, and the almost complex structure ${\tilde
J}_{ab}$ is defined in \eq{jtilde}.

These are remarkable solutions in that they are, to our best
knowledge, the first exact solution of higher spin gauge theory in
which higher spin fields are non-vanishing.  We also note that the
Weyl tensors in these solutions corresponds to higher spin
generalization of the Type D Weyl tensor that takes the form
$\phi_{\ad\bd\cd\dd} \sim \lambda_{(\ad}\lambda_{\bd} \mu_{\cd}
\mu_{\dd)}$ up to a scale factor \cite{flaherty}. Type D instanton
solutions of Einstein's equation in Euclidean signature with and
without cosmological constant have been discussed in
\cite{Lapedes:1980qw}. Our solution provides their higher spin
generalization.

After we describe the HS field equations in diverse signatures in
Section 2, we shall present the detailed construction of our
solutions in Section 3. We shall comment further on these solutions
and open problems in the Conclusions.


\scs{The Bosonic 4D Models in Various Signatures}


We shall first describe the field equations without imposing reality
conditions on the master fields. These conditions will then be
discussed separately leading to five different models in
four-dimensional spacetimes with various signatures (see Table
\ref{Table1}).


\scss{The Complex Field Equations}


To formulate the complex field equations we use independent
$SL(2;\Comp)_L$ doublet spinors $(y_\a,z_\a)$ and $SL(2;\Comp)_R$
doublet spinors $(\yb_{\ad},\zb_{\ad})$ generating an oscillator
algebra with non-commutative and associative product $\star$ defined
by
\bea y_\a\star y_\b&=&y_\a y_\b+i\epsilon_{\a\b}\ ,\qquad
y_{\a}\star z_{\b}\ =\ y_{\a}z_{\b}-i\,\e_{\a\b}\ ,\label{osc1}\\[5pt] z_{\a}\star
y_{\b}&=& z_{\a}y_{\b}+i\,\e_{\a\b}\ , \qquad z_{\a}\star z_{\b}\ =\
z_{\a}z_{\b}-i\,\e_{\a\b} \ ,\label{osc2} \eea
and
\bea \bar y_{\dot\a}\star \bar y_{\dot\b}\ =\ \bar y_{\dot\a} \bar
y_{\dot\b}+i\epsilon_{\dot\a\dot\b}\ ,\qquad \bar z_{\dot\a}\star
\bar y_{\dot\b}\ =\ \bar z_{\dot \a} \bar y_{\dot\b}-
i\epsilon_{\dot\a\dot\b}\ ,\label{oscbar1}\\[5pt] \bar y_{\dot\a}\star \bar z_{\dot\b}\
=\ \bar y_{\dot\a} \bar z_{\dot\b}+i\epsilon_{\dot\a\dot\b}\ ,\qquad
\bar z_{\dot\a}\star \bar z_{\dot\b}\ =\ \bar z_{\dot\a} \bar
z_{\dot\b}-i\epsilon_{\dot\a\dot\b}\ ,\label{oscbar2}\eea
where the juxtaposition denotes the symmetrized, or Weyl-ordered,
products. For example, $y_\a y_\b=\ft12(y_\a\star y_\b+y_\b\star
y_\a)$. Equivalently, Weyl-ordered functions obey\footnote{The
integration measure is defined by $d^4\xi=d^2\x^1 d^2\x^2$, where
$d^2 z=idz\wedge d\bar z=2dx\wedge dy$ for $z=x+iy$. With this
normalization, $\id\star \widehat f=\widehat f$.}
 \bea
 &&\widehat f(y,\bar y,z,\bar z)~\star~ \widehat g(y,\bar
y,z,\bar z)\label{star}\\[5pt]&=&\ \int \frac{d^4\xi d^4\eta}{(2\pi)^4}~ e^{i\eta^\a\xi_\a+
i\bar\eta^{\dot\a}\bar\x_{\dot\a}} ~\widehat f(y+\xi,\bar y+\bar
\xi,z+\xi,\bar z-\bar \xi)~\widehat g(y+\eta,\bar y+\bar
\eta,z-\eta,\bar z+\bar \eta)\ ,\nn
 \eea
where the hats are used to denote functions of all oscillators,
while functions of only $y_\a$ and $\yb_{\ad}$ will be unhatted.

The complex master fields are the \emph{adjoint} one-form $\widehat
A$ and the \emph{twisted-adjoint} zero-form $\widehat \Phi$ defined
by
\bea \widehat A&=&dx^\mu\widehat A_\mu(x;y,\bar y,z,\bar
z)+dz^\a\widehat A_\a(x;y,\bar y,z,\bar z)+d\bar z^{\dot\a}\widehat
A_{ \dot\a}(x;y,\bar y,z,\bar z)\ ,\qquad\\[5pt]
\widehat\Phi&=&\widehat \Phi(x;y,\bar y,z,\bar z)\ ,\eea
where $x^\mu$ are coordinates on a commutative base manifold (which
can, but need not, be fixed to be four-dimensional spacetime). One
also defines the total exterior derivative
\be d\ =\ dx^\m\partial_\m + dz^\a {\partial\over\partial z^{\a}}+
d\bar z^{\ad} {\partial\over\partial \bar z^{\ad}}\ ,\ee
with the property $d(\widehat f\wedge\star~\widehat g)=(d\widehat
f)\wedge \star~\widehat g+(-1)^{{\rm deg}\widehat f}\widehat
f\wedge\star ~d\widehat g$ for general differential forms. In what
follows we shall suppress the $\wedge$. The master fields can be
made subject to the following discrete symmetry
conditions\footnote{The exterior derivative obeys $\tau d=d\tau$ and
$\pi d=d\pi$, and the $\tau$ and $\pi$ maps do not act on the
commutative coordinates.} \cite{Vasiliev:1995dn,Engquist:2002vr}
\bea \mbox{Minimal model ($s=0,2,4,...$)}&:& \tau(\widehat A)\ =\
-\widehat A\
,\qquad \tau(\widehat\Phi)\ =\ \bar\pi(\widehat \Phi)\ ,\label{minmod}\\[5pt]
\mbox{Non-minimal model ($s=0,1,2,3,...$)}&:& \pi\bar\pi(\widehat
A)\ =\ \widehat A\ ,\qquad \pi\bar\pi(\widehat \Phi)\ =\
\widehat\Phi\ ,\label{nonminmod}\eea
where $\tau$ is the $\star$-product algebra anti-automorphism
defined by
\be \tau(\widehat f(y,\bar y;z,\bar z))\ =\ \widehat f(iy,i\bar
y;-iz,-i\bar z)\ ,\label{tau}\ee
and $\pi$ and $\bar\pi$ are two involutive $\star$-product
automorphisms defined by
\bea \pi(\widehat f(y,\bar y;z,\bar z))&=&\widehat f(-y,\bar
y;-z,\bar z)\ ,\qquad \bar\pi(\widehat f(y,\bar y;z,\bar z))\ =\
\widehat f(y,-\bar y;z,-\bar z)\ .\eea
We note that
\bea \tau(\widehat f\star\widehat g)&=&(-1)^{{\rm deg}(\widehat
f){\rm
deg}(\widehat g)}\tau(\widehat g)\star\tau(\widehat f)\ ,\\[5pt]
\pi(\widehat f\star\widehat g)&=& \pi(\widehat f)\star\pi(\widehat
g)\ ,\\[5pt] \bar\pi(\widehat f\star\widehat g)&=& \bar\pi(\widehat
f)\star\bar\pi(\widehat g)\ ,\eea
and that $\tau^2=\pi\bar\pi$. The automorphisms are inner and can be
generated by conjugation with the functions $\kappa$ and
${\bar\kappa}$ given by
 \be
\k\ =\ \exp(iy^\a z_\a)\ ,\quad\quad \bar{\k}\ =\ \exp(-i\bar
y^{\ad}\bar z_{\ad})\ ,
 \ee
such that
\be \kappa\star \widehat f(y,z)\ =\ \kappa \widehat f(z,y)\ ,\qquad
\widehat f(y,z)\star\kappa\ =\ \kappa \widehat f(-z,-y)\ ,\qquad
\kappa\star \widehat f\star\kappa\ =\ \pi(\widehat f)\ ,
\label{kappa}\ee \be \bar \kappa\star \widehat f(\bar y,\bar z)\ =\
\bar\kappa \widehat f(-\bar z,-\bar y)\ ,\qquad \widehat f(\bar
y,\bar z)\star\bar \kappa\ =\ \bar\kappa \widehat f(\bar z,\bar y)\
,\qquad \bar\kappa\star \widehat f\star\bar\kappa\ =\
\bar\pi(\widehat f)\ . \label{kappabar}\ee
The full complex field equations are
\bea \widehat F&=& \frac{i}4 \left[c_1 dz^\a\wedge dz_\a \widehat
\Phi\star \kappa+c_2 d\bar z^{\ad}\wedge d\bar z_{\ad}\widehat
\Phi\star\bar\kappa\right]\ ,\label{m1}\\[5pt]\widehat D\widehat \Phi&=&0\
,\label{m2}\eea
where $c_1$ and $c_2$ are complex constants and the curvatures and
gauge transformations are given by
\bea \widehat{F}&=& d\widehat{A}+\widehat{A}\star\widehat{A}\
,\qquad\ \delta_{\widehat \e}\widehat A\ =\ \widehat D\widehat
\e\\[5pt]\widehat{D}\widehat{\Phi}&=&d\widehat{\Phi}+[\widehat
A,\widehat \Phi]_\pi\ ,\qquad\delta_{\widehat \e}\widehat\Phi\ =\
-[\widehat \e,\widehat\Phi]_\pi\ ,\eea
with
\bea [\widehat f,\widehat g]_\pi&=&\widehat f\star\widehat
g-(-1)^{{\rm deg}(\widehat f){\rm deg}(\widehat g)}\widehat
g\star\pi(\widehat f)\ .\eea
Since $\widehat \Phi$ is defined up to rescalings by complex
numbers, the model only depends on one complex parameter, that we
can take to be
\bea c&=& {c_2\over c_1}\ .\eea
In components, the constraints read
\bea \widehat F_{\m\n}&=&0\ ,\qquad \widehat D_\mu\widehat\Phi\
\equiv\ \partial_\mu\widehat\Phi +[\widehat
A_\m,\widehat\Phi]_{\pi}\ =\ 0\ , \label{f1}\eea
\bea \widehat F_{\m\a}&=&0\ ,\qquad \widehat F_{\m\ad}\ =\ 0\
,\label{f2}\eea
\bea \widehat
F_{\a\b}&=&-\ft{ic_1}2\e_{\a\b}\widehat\Phi\star\kappa\ ,\qquad
\widehat F_{\ad\bd}\ =\
-\ft{ic_2}2\e_{\ad\bd}\widehat\Phi\star\bar\kappa\ ,\label{f3}\eea
\bea \widehat F_{\a\ad}&=&0\ ,\label{f4}\eea
\bea \widehat D_\a\widehat\Phi\ \equiv\
\partial_\a\widehat\Phi+\widehat
A_\a\star\widehat\Phi+\widehat\Phi\star\pi(\widehat A_\a)&=&0\
,\label{s1}\eea
\bea \widehat D_{\ad}\widehat\Phi\ \equiv\
\partial_{\ad}\widehat\Phi+\widehat
A_{\ad}\star\widehat\Phi+\widehat\Phi\star\bar\pi(\widehat
A_{\ad})&=&0\ ,\label{s2}\eea
where \eq{s2} can be derived using $\pi\bar\pi(\widehat
A_{\ad})=-\widehat A_{\ad}$. Introducing \cite{vasiliev}
\bea \widehat S_\a&=& z_\a-2i\widehat A_\a\ ,\qquad \widehat
S_{\ad}\ =\ \bar z_{\ad}-2i\widehat A_{\ad}\ ,\label{S}\eea
the component form of the equations carrying at least one spinor
index now take the form
\bea \partial_\m \widehat S_\a+[\widehat A_\mu,\widehat
S_\a]_\star&=&0\ ,\qquad
\partial_\m \widehat
S_{\ad}+[\widehat A_\mu,\widehat S_{\ad}]_\star\ =\ 0\
,\label{S1}\eea \bea [\widehat S_\a,\widehat S_\b]_\star&=&
-2i\e_{\a\b}(1-c_1\widehat \Phi\star\kappa)\ ,\qquad[\widehat
S_{\ad},\widehat S_{\bd}]_\star\ =\  -2i\e_{\ad\bd}(1-c_2\widehat
\Phi\star\bar\kappa)\ ,\label{S2}\eea\bea [\widehat S_\a,\widehat
S_{\bd}]_\star&=&0\ ,\label{S3}\eea \bea \widehat
S_\a\star\widehat\Phi+\widehat\Phi\star\pi(\widehat S_\a)&=&0\
,\label{S4}\eea\bea \widehat
S_{\ad}\star\widehat\Phi+\widehat\Phi\star\bar\pi(\widehat S_{\ad})\
=\ 0\ .\label{S5}\eea
This form of the equations makes the following $\integ_2\times
\integ_2$ symmetry manifest:
\bea \widehat S_\a&\rightarrow\pm \widehat S_\a\ ,\qquad \widehat
S_{\ad}\ \rightarrow\ \pm\widehat S_{\ad}\ ,\label{Z2}\eea
(where the two transformations can be performed independently)
keeping $\widehat A_\m$ and $\widehat \Phi$ fixed. We note that
$\widehat S_\a\rightarrow -\widehat S_\a$ is equivalent to $\widehat
A_\a\rightarrow-\widehat A_\a-i z_\a$, \emph{idem} $\widehat
S_{\ad}$ and $\widehat A_{\ad}$.

All component fields are of course complex at this level. Next we
shall discuss various reality conditions on the (hatted) master
fields that will lead to models with real physical fields living in
spacetimes with different signatures.


\scss{Real Forms}


In order to define the real forms of the field equations one has to
impose reality conditions on both adjoint one-form and
twisted-adjoint zero-form, corresponding to suitable real forms of
the higher-spin algebra and signatures of spacetime. There are three
distinct real forms of the complex higher-spin algebra itself. In
two of these cases there are two distinct reality conditions that
can be imposed on the zero-form, leading to five distinct models in
total, as shown in Table \ref{Table1}. The reality conditions are
\be \widehat A^\dagger\ =\ -\sigma(\widehat A)\ ,\qquad
\widehat\Phi^{\dagger}\ =\ \sigma(\pi(\widehat \Phi))\
,\label{dagger}\ee
where the possible actions of the dagger
\footnote{The dagger acts as usual complex conjugation on component
fields; in this paper we shall denote the conjugate of a complex
number $x$ by $x^\ast$, while reserving the bar for denoting
quantities associated with the $R$-handed oscillators.} on the
spinor oscillators and consequential selections of real forms of
$SO(4;\Comp)\simeq SL(2;\Comp)\times SL(2;\Comp)$ are given by
\bea SU(2)_L\times SU(2)_R&:&\quad (y^\a)^\dagger\ =\
y^{\dagger}_{\a}\ ,\quad (z^\a)^\dagger\ =\ z^{\dagger}_{\a}\
,\label{su2}\\&&\quad  (\bar y^{\ad})^\dagger\ =\ \bar
y^{\dagger}_{\ad}\ ,\quad (\bar
z^{\ad})^\dagger\ =\ \bar z^{\dagger}_{\ad}\ ,\nn\\[5pt]
SL(2;\Comp)_{\rm diag}&:&\quad (y^\a)^\dagger\ =\ \bar y^{\ad}\
,\quad
(z^\a)^\dagger\ =\ \bar z^{\ad}\ ,\label{sl2}\\[5pt]
Sp(2;\Real)_L\times Sp(2;\Real)_R&:&\quad (y^\a)^\dagger\ =\ y^{\a}\
,\quad (z^\a)^\dagger\ =\ -z^\a\ ,\label{sp2}\\&&\quad (\bar
y^{\ad})^\dagger\ =\ \bar y^{\ad}\ ,\quad (\bar z^{\ad})^\dagger\ =\
-\bar z^{\ad} \ ,\nn
\eea
and the map $\sigma$ is given in Table \ref{Table1}, with the
isomorphism $\rho$ given by
\be \rho(\widehat f(y^\dagger_\a,\bar y^\dagger_\a,z^\dagger_\a,\bar
z^\dagger_\a))\ =\ \widehat f(y_\a,\bar y_\a,-z_\a,-\bar z_\a)
\label{iso}\ee
in the case of $(4,0)$ signature. Note that $\sigma$ is an
oscillator-algebra automorphism in signatures $(3,1)$ and $(2,2)$,
while it is an isomorphism in signature $(4,0)$. Here, the $SU(2)$
doublets are pseudo real in the sense that from
$(y_\a)^\dagger=-y^{\dagger\a}$ \emph{idem} $(z_\a)^\dagger$, $(\bar
y_{\ad})^\dagger$ and $(\bar z_{\ad})^\dagger$ it follows that
$(y_\a,\yb_{\ad};z_\a,\zb_{\ad})$ and
$(y^\dagger_\a,\yb^\dagger_\a;z^\dagger_\a,\zb^\dagger_{\ad})$
generate equivalent oscillator algebras with isomorphism $\rho$. The
reality property of the exterior derivative takes the following form
in different signatures:
\bea \mbox{Signature $(3,1)$ and $(2,2)$}&:& \quad d^\dagger\ =\ d\
,\\\mbox{Signature $(4,0)$}&:& \quad \rho \circ d^\dagger\ =\ d\circ
\rho\ .\eea
We note that the Euclidean case is consistent in the sense that
\bea \rho (dz^\a)^\dagger&=&\rho d^\dagger (z^\a)^\dagger\ =\ d\rho
(z^\dagger_\a)\ =\ -dz_\a\eea
is compatible with representing $d\widehat f$ using
$\partial\widehat f/\partial z^\a=\frac i2 [z_\a,\widehat f]_\star$,
which yields
\bea \rho\left(  \frac i2 dz^\a [z_\a,\widehat
f]_\star\right)^\dagger&=& \frac i2 dz_\a \rho\left([\widehat
f^\dagger,-z^{\dagger\a}]_\star\right)\ =\ \frac i2
dz_\a[\rho\widehat
 f^\dagger,z^{\a}]_\star\ =\ \frac i2 dz^\a [z_\a,\rho\widehat
 f^\dagger]_\star\ .\eea
Demanding compatibility between the reality conditions \eq{dagger}
and the master field equations \eq{m1} and \eq{m2}, and using
\bea \rho\left((\kappa)^\dagger\right)&=&\kappa\ ,\qquad
\rho\left((idz^\a\wedge dz_\a)^\dagger\right)\ =\ -idz^\a\wedge
dz_\a\ ,\eea
one finds the following reality conditions on the parameters
\bea \mbox{Signature $(3,1)$}&:& c_1^\ast\ =\ c_2\ ,\\
\mbox{Signature $(4,0)$ and $(2,2)$}&:& c_1^\ast\ =\ c_1\ ,\quad
c_2^\ast\ =\ c_2\ .\eea
As a result, the parameter $c$ is a phase factor in Lorentzian
signature and a real number in Euclidean and Kleinian signatures.
The parameters can be restricted further by requiring invariance
under the parity transformation
\bea P(y_\a)&=& \yb_{\ad}\ ,\qquad P\circ d\ =\ d\circ P\ ,\qquad
P^2\ =\ {\rm Id}\ .\eea
Taking $\widehat A$ to be invariant and assigning intrinsic parity
$\e=\pm 1$ to $\widehat\Phi$,
\bea P(\widehat A)&=&\widehat A\ ,\qquad P(\widehat\Phi)\ =\ \e
\widehat\Phi\ ,\eea
one finds that the master equations are parity invariant provided
that \cite{Sezgin:2003pt}
\bea c&=&\e\ =\ \mx{\{}{ll}{1&\mbox{Type A model (scalar)}\\[5pt]
-1&\mbox{Type B model (pseudoscalar)}}{.}\eea
In Lorentzian signature, there is no loss of generality in choosing
$c_1=c_2=1$ in the Type A model and $c_1=-c_2=i$ in the Type B
model, while in Euclidean and Kleinian signatures, one may always
take $c_1=c_2=1$ in the Type A model and $c_1=-c_2=1$ in the Type B
model. More generally, the parity transformation maps different
models into each other as follows,
\bea P(c_1)&=&\e c_2\ ,\qquad P(c_2)\ =\ \e c_1\ ,\qquad P(c)\ =\
{1\over c}\ ,\eea
leaving invariant the Type A and B models. The \emph{maximally
parity violating} cases are
\bea \mbox{Signature $(3,1)$}&:& c\ =\ \exp(i\pi/4)\ ,\\[5pt]
\mbox{Signature $(4,0)$ and $(2,2)$}&:& c\ =\ 0\ .\eea
The case with $c=0$ shall be referred to as the \emph{chiral model},
that we shall discuss in more detail below.

The HS equations in Lorentzian signature have the $\integ_2$
symmetry acting as $({\widehat S}_\a, {\widehat S}_{\ad})
\rightarrow (\epsilon {\widehat S}_\a, \epsilon {\widehat
S}_{\ad})$, and $\integ_2\times \integ_2$ symmetry in $(4,0)$ and
$(2,2)$ signatures acting as $({\widehat S}_\a, {\widehat S}_{\ad})
\rightarrow (\epsilon {\widehat S}_\a, \epsilon' {\widehat
S}_{\ad})$, where $\epsilon=\pm 1$ and $\e'=\pm 1$.


Finally, let us give the reality conditions at the level of the
$SO(5;\Comp)$ algebra and its minimal bosonic higher-spin extension.
The adjoint representation of the complex minimal bosonic
higher-spin Lie algebra is defined by
\footnote{A more detailed description of the complex higher-spin
algebra and its representations is given in \cite{wip}.}
\bea \mhso(5;\Comp) &=& \left\{Q(y,\bar y)\ :\quad \tau(Q)\ =\
-Q\right\} \ ,\label{chsa}\eea
and the corresponding minimal twisted-adjoint representation by
\bea T[\mhso(5;\Comp)] &=& \left\{S(y,\yb)\ :\quad \quad \tau(S)\ =\
\pi(S)\right\}\ .\label{ctwadj}\eea
The real forms are defined by
\bea\mhso(5-q,q) &=& \left\{Q(y,\bar y)\in \mhso(5;\Comp)\ :\quad
Q^\dagger\ =\ -\sigma(Q)\right\} \ ,\label{chsareal}\\[5pt]
T[\mhso(5-q,q)] &=& \left\{S(y,\yb)\in T[\mhso(5-q,q)\ :\quad
S^\dagger\ =\ \sigma(\pi(S))\right\}\ .\label{ctwadj}\eea
The finite-dimensional $SO(5;\Comp)$ subalgebra is generated by
$M_{AB}$, that we split into Lorentz rotations and translations
$(M_{ab},P_a)$ defined by
\bea \pi(M_{ab})&=&M_{ab}\ ,\qquad \pi(P_a)\ =\ -P_a\ .\eea
For these generators, which by convention arise in the expansion of
the master fields together with a factor of $i$, the reality
condition \eq{dagger} implies
\bea (M_{AB})^\dagger&=&\sigma(M_{AB})\ .\eea
This condition is solved by
\be M_{ab}\ =\ -\frac18\left((\s_{ab})^{\a\b}y_\a
y_\b+(\bar\sigma_{ab})^{\ad\bd}\yb_{\ad}\yb_{\bd}\right),\quad P_a\
=\ \frac{\l}4(\sigma_a)^{\a\ad}y_\a\yb_{\ad}\ ,\ee
where the van der Waerden symbols are defined in Appendix \ref{App1}
and $\l^2$ is proportional to the cosmological constant, as shown in
Table \ref{Table1}. The van der Waerden symbols encode the spacetime
signature $\eta_{ab}$, and the commutation relations among the
$M_{AB}$ then fix the signature of the ambient space to be
\bea \eta_{AB}&=&(\eta_{ab};-\lambda^2)\ .\eea
%


\scss{The Chiral Model}


In the chiral model with $c=0$, the master field $\widehat\Phi$ can
be eliminated using \eq{f3}, and expressed as
\bea \widehat \Phi&=&(1+\frac{i}2 \widehat S^\a\star\widehat
S_\a)\star\kappa\ ,\eea
where we have chosen $c_1=1$ and $\widehat S_\a$ is given by \eq{S}.
The remaining independent master-field equations now read
\bea \widehat F_{\m\n}&=&0\ ,\qquad \widehat D_\m\widehat S_\a\ =\
0\ ,\qquad \widehat D_\m\widehat S_{\ad}\ =\ 0\ ,\eea\bea[\widehat
S_\a,\widehat S_{\ad}]_\star&=&0\ ,\qquad [\widehat S_{\ad},\widehat
S_{\bd}]_\star\ =\ -2i\e_{\ad\bd}\ ,\eea\bea \widehat S_\a\star
\widehat S^\b\star\widehat S_\b+\widehat S^\b\star\widehat S_\b\star
\widehat S_\a&=&4i \widehat S_\a\ .\label{SplusScube}\eea
We note that \eq{S5} holds identically in virtue of $\widehat
S_{\ad}\star\widehat\Phi+\widehat\Phi\star\bar\pi(\widehat
S_{\ad}))=[\widehat S_{\ad},1+\frac i2 \widehat S^\a\star\widehat
S_\a]_\star\star\kappa=0$, where we used $\kappa\bar\kappa\star
\widehat S_{\ad}\star\kappa\bar\kappa=-\widehat S_{\ad}$ and
$[\widehat S_\a,\widehat S_{\ad}]_\star=0$. The chiral model can be
truncated further by imposing
\bea \widehat A_{\ad}&=&0\ ,\qquad {\partial\over \partial
z^{\ad}}\widehat A_\mu\ =\ 0\ ,\qquad {\partial\over \partial
z^{\ad}}\widehat A_\a\ =\ 0\ .\eea
In general, the chiral model also has interesting solutions with
non-vanishing $\widehat A_{\ad}$, since flat connections in
non-commutative geometry can be non-trivial.


\scss{Comments on the Weak-Field Expansion and Spectrum}


The procedure, described in great detail in \cite{Sezgin:2002ru},
for obtaining the manifestly diffeomorphism and locally Lorentz
invariant weak-field expansion of the physical field equations can
be extended straightforwardly to arbitrary signature. The expansion
is in terms of spin-$s$ physical fields with $s\neq 2$ as well as
higher derivatives of all fields, while the vierbein and Lorentz
connection are treated exactly.

In this approach one first solves \eq{f2}--\eq{s2} subject to the
initial condition
\bea \Phi &=& \widehat\Phi|_{Z=0}\ ,\\[5pt]
A_\mu&=& \left.\widehat A_\m\right|_{Z=0}\ =\
e_\mu+\omega_\mu+W_\m+K_{\m}\ ,\label{Amu}\eea
where
\bea e_\m&=&{1\over 2i}e_\m{}^a P_a\ ,\qquad \o_\m \ =\ \ft1{4i}
\o_\m{}^{ab}M_{ab}\ ;\label{em}\eea
$W_\m$ contains the higher-spin gauge fields (and also the spin
$s=1$ gauge field in the non-minimal model); and the field
redefinition
\bea K_{\m}&=& {1\over 4i}\o_\m{}^{\a\b}\left.\widehat
S_\a\star\widehat S_\b\right|_{Z=0}+{1\over 4i}
\bar\omega_\m{}^{\ad\bd}\left.\widehat S_{\ad}\star\widehat
S_{\bd}\right|_{Z=0}\\[5pt]&=& i\o_{\m}{}^{\a\b} \left.(\widehat
A_\a\star\widehat A_\b-{\partial\over\partial y^\a} \widehat
A_\b)\right|_{Z=0}+ i\bar\o_{\m}{}^{\ad\bd} \left.(\widehat
A_{\ad}\star\widehat A_{\bd}-{\partial\over\partial \yb^{\ad}}
\widehat A_{\bd})\right|_{Z=0}\ .\label{Kmu}\eea
One also imposes the gauge condition
\bea \widehat A^{(0)}_{\a}&=&0\ ,\qquad \widehat A^{(0)}_{\ad}\ =\
0\ ,\label{gauge}\eea
where we have defined the internal flat connection
\bea \widehat A^{(0)}_{\a}&=&\widehat A_\a|_{\Phi=0}\ ,\qquad
\widehat A^{(0)}_{\ad}\ =\ \widehat A_{\ad}|_{\Phi=0}\ .\eea
One then substitutes the resulting $\widehat\Phi$ and $\widehat
A_\mu$, which can be obtained explicitly in a perturbative expansion
in $\Phi$, into \eq{f1} and sets $Z=0$, which yields a manifestly
spin-2 covariant complex HS gauge theory on the base manifold. Up to
this point the local structure of the base-manifold, nor the
detailed structure of the gauge fields, have played any role. To
proceed, one may refer to an ordinary spacetime, take $e_\mu{}^a$ to
be an (invertible) vierbein, and treat $W_\mu$ as a weak field. This
allows one to eliminate a large number of auxiliary fields in $\Phi$
and $W_\mu$, leaving a model consisting of a physical scalar
$\phi=\Phi|_{y=\bar y=0}$, the vierbein $e_\mu{}^a$, and an infinite
tower of (doubly traceless) HS gauge fields $\phi_{a(s)}$ residing
in $W_\mu$.

The gauge choice \eq{gauge} is convenient since it implies
${\partial\over\partial y^\a} \widehat A_\b|_{Z=0}=0$ that
simplifies the expansion \cite{Sezgin:2002ru}. However, there are
also other gauges where $\widehat A_\a|_{\Phi=0}$ is a flat but
non-trivial internal connection, and indeed this will be the case
for the Type 1 and Type 2 solutions that we shall present in Section
3.

In the leading order in the weak fields, the two-form and one-form
constraints for the minimal model read
\bea \mbox{$s=2$} &:& \left\{ \ba{ll} {\cal R}_{\a\b,\c\d}\ =\
c_2\Phi_{\a\b\c\d}\ ,\qquad&{\cal R}_{\ad\bd,\c\d}\ =\ 0\ ,\w2 {\cal
R}_{\a\b,\c\dd}\ =\ 0\ ,&{\cal R}_{\ad\bd,\c\dd}\ =\ 0\ ,\w2 {\cal
R}_{\a\b,\cd\dd}\ =\ 0\ ,&{\cal R}_{\ad\bd,\cd\dd}\ =\
c_1\Phi_{\ad\bd\cd\dd}\ ,\ea \right. \label{s2b} \w4
\mbox{$s=4,6,...$}&:& \left\{ \ba{ll} F^{(1)}_{\a\b,\c_1\dots
\c_{2s-2}}\ =\ c_2\Phi_{\a\b\c_1\dots \c_{2s-2}}\
,&F^{(1)}_{\ad\bd,\cd_1\dots\cd_k\c_{k+1}\dots \c_{2s-2}}\ =\  0\
,\la{hse1}\w3 F^{(1)}_{\a\b,\c_1\dots\c_k\cd_{k+1}\dots \cd_{2s-2}}\
=\ 0\ ,&F^{(1)}_{\ad\bd,\cd_1\dots \cd_{2s-2}}\ =\
c_1\Phi_{\ad\bd\cd_1\dots \cd_{2s-2}}\ ,\ea\right. \label{hs} \w4
\mbox{0-forms}&:& \nabla_{\a}{}^{\ad}\Phi_{\b_1\dots\b_m}{}^{
\bd_1\dots\bd_n}\ =\
i\lambda\left(\Phi_{\a\b_1\dots\b_m}{}^{\ad\bd_1\dots\bd_n}-
 mn\e_{\a(\b_1}\e^{\ad(\bd_1}\Phi_{\b_2\dots
\b_m)}{}^{\bd_2\dots\bd_n)}\right)\ ,\phantom{aaaaa}
\la{linscalareq}\eea
where for higher spins $s=4,6,\dots$ and $k=0,\dots,2s-3$, and for
$0$-forms $|m-n|=0$ mod $4$. In all cases, the zero-form system
contains a physical scalar with field equation
\bea (\nabla^2+2\lambda^2)\phi&=&0\ .\eea
In the Lorentzian case, where both $c_1$ and $c_2=c_1^\ast$ are
non-zero, the spin-2 sector consists of gravity with cosmological
constant $-3\lambda^2$, and the spin-$s$ sectors with $s=4,6,\dots$
consist of higher-spin tensor gauge fields with critical masses
proportional to $\lambda^2$. The criticality in the masses, that
implies composite masslessness\footnote{By definition masslessness
refers to reduction in the infinite-dimensional weight space of the
various real forms of $SO(5;\Comp)$. This is well-known for
$SO(3,2)$ and similar situations arise for other signatures as well.
By compositeness we mean that the massless states are composites of
singletons \cite{Flato:1978qz}.} in the case of AdS, holds in the dS
case as well, where thus the physical spectrum is given by the
symmetric tensor product of two (non-unitary) $SO(4,1)$ singletons
\cite{wip}.

In the Euclidean and Kleinian cases, the parameters $c_1$ and $c_2$
are real and independent. In case $c_1c_2\neq 0$, the Lorentzian
analysis carries over, leading to a composite massless spectrum
given by symmetric tensor products of suitable singletons
\cite{wip}. However, unlike the Lorentzian case, the spin-$s$ sector
of the twisted adjoint representation can be decomposed into
left-handed and right-handed sub-sectors of real states,
corresponding to $\{\Phi_{\a_1\dots\a_m,\ad_1\dots\ad_n}\}$ with
$m-n=\pm 2s$ \cite{wip}. These sub-sectors mix under HS
transformations.

In case either $c_1$ or $c_2$, but not both, vanishes, that we shall
refer to as the chiral models, the metric and the higher-spin gauge
fields become half-flat. For definiteness, let us consider the case
$c_2=0$. The components of the zero-form that drop out in the
two-form constraint, \emph{i.e.} $\Phi_{\a_1\dots\a_{2s}}$, now
become \emph{independent} physical fields, obeying field equations
following from \eq{linscalareq}.

{\footnotesize \tabcolsep=1mm \begin{table}[h!]
\begin{tabular}{|c|c|c|c|c|c|c|}\hline
& & & & & & \\
HSA & Signature & Spinors &  &
Reality & Symmetric & Hermitian \\
&  &  & $\mbox{\phantom{aaaaa}}$ &
  &  space&isometries  \\
 & \mbox{\footnotesize $\eta_{ab}$} &  &$\lambda^2$& $\sigma$ & &
 \\ \hline & & & & & & \\
\mbox{\footnotesize $\mhso(5)$} & \mbox{\footnotesize $(4,0)$} &
\mbox{\footnotesize $SU(2)_L\times SU(2)_R$} & $-1$ & $\rho$ &
\mbox{\footnotesize $S^4$} & \mbox{\footnotesize $\mso(2) \otimes
\mso(3)$}
\\ \mbox{\footnotesize $\mhso(4,1)$} & \mbox{\footnotesize $(4,0)$}
& \mbox{\footnotesize $SU(2)_L\times SU(2)_R$} & $+1$ & $\rho\pi$
& \mbox{\footnotesize $H_4$} & \mbox{\footnotesize $\mso(3,1)$} \\
\mbox{\footnotesize $\mhso(4,1)$} & \mbox{\footnotesize $(3,1)$} &
\mbox{\footnotesize $SL(2,\Comp)_{\rm diag}$} & $-1$ & $\pi$ &
\mbox{\footnotesize $dS_4$} &
\mbox{\footnotesize $\mso(3,1)'$} \\
\mbox{\footnotesize $\mhso(3,2)$} & \mbox{\footnotesize $(3,1)$} &
\mbox{\footnotesize $SL(2,\Comp)_{\rm diag}$} & $+1$ & id &
\mbox{\footnotesize $AdS_4$} &
\mbox{\footnotesize $\mso(3,2)$} \\
\mbox{\footnotesize $\mhso(3,2)$} & \mbox{\footnotesize $(2,2)$} &
\mbox{\footnotesize $SL(2,\Real)_L\times SL(2,\Real)_R$} & $-1$ & id
& \mbox{\footnotesize $H_{3,2}$} & \mbox{\footnotesize $\mso(3,2)$}
\\ & & & & & & \\
\hline
\end{tabular}
\caption{{\footnotesize The minimal bosonic higher-spin algebras
$\mhso(p',5-p')\supset \mso(5-p',p')$ in signature $(p,4-p)$ can be
realized with spinor oscillators transforming as doublets under the
groups listed in the third column. These realizations obey reality
conditions $(M_{AB})^\dagger=\sigma(M_{AB})$, with hermitian
subalgebras listed above \cite{wip}. The symmetric spaces with unit
radius have cosmological constant $\Lambda=-3\lambda^2$.}}
\label{Table1}
\end{table}}


\scs{Exact Solutions}


In this section we shall give four types of exact solutions to the
4D HS models given in the previous section. The salient features of
these are summarized in the Introduction. Here we stress that (a)
the Type 0 solutions are maximally symmetric spaces; (b) the Type 1
solutions are $SO(4-p,p)$ invariant deformations of Type 0; (c) the
Type 2 solutions, which exist necessarily in the non-minimal model,
have vanishing spacetime component fields but non-vanishing
spinorial master one-form; (d) the Type 3 solutions, which exist in
the non-minimal chiral model only, have the remarkable feature that
all higher spin gauge fields are non-vanishing in such a way that
the Weyl zero-forms are covariantly constant, in a certain sense
that will be explained below. Before we give these four types of
solutions we shall describe briefly the method for solving the
master field equations using gauge functions.


\scss{The Gauge Function Ansatz}


In order to construct an interesting class of solutions we shall use
the $Z$-space approach \cite{Bolotin:1999fa,Sezgin:2005pv} in which
the constraints carrying at least one curved spacetime index,
\emph{viz.}
 \bea
 \widehat F_{\m\n}&=&0\ ,\qquad \widehat D_\m \widehat\Phi\ =\ 0\ ,\\[5pt]
 \widehat F_{\m\a}&=&0\ ,\qquad
 \widehat F_{\m\ad}\ =\ 0 \ ,\label{xsp2}
 \eea
are integrated in simply connected spacetime regions given the
spacetime zero-forms at a point $p$,
\be
 \widehat \Phi'\ =\ \widehat
\Phi|_{p}\ ,\qquad \widehat S'_\a\ =\  \widehat S_\a|_{p}\ ,\qquad
\widehat S'_{\ad}\ =\ \widehat S_{\ad}|_{p}\ , \label{phiprime}
 \ee
and expressed explicitly as
 \bea
\widehat A_\mu&=&\widehat L^{-1}\star \partial_\mu \widehat L\
,\qquad \widehat \Phi\ =\ {\widehat L}^{-1}\star \widehat\Phi'\star
\pi(\widehat L)\ ,\qquad\\[5pt]\widehat S_\a&=&\widehat L^{-1}\star
\widehat S'_\a\star \widehat L\ ,\qquad \widehat S_{\ad}\ =\
\widehat L^{-1}\star \widehat S'_{\ad}\star \widehat L\ ,
\label{Leq}
 \eea
where $\widehat L=\widehat L(x,z,\bar z;y,\yb)$ is a gauge function,
and
\be \widehat L|_{p}\ =\ 1\ ,\qquad
\partial_\m\widehat\Phi'\ =\ 0\ ,\qquad \partial_\m{\widehat S}'_\a\ =\
0\ ,\qquad \partial_\m{\widehat S}'_{\ad}\ =\ 0\ .\ee
The internal connections $\widehat A_\a$ and $\widehat A_{\ad}$ can
be reconstructed from $\widehat S_\a$ and $\widehat S_{\ad}$ using
\eq{S}. In particular note the relation
\be {\widehat A}_\a = {\widehat L}\star \partial_\a {\widehat L}
+{\widehat L}^{-1}\star {\widehat A}'_\a \star {\widehat L}\
,\label{Aalpha} \ee
and it follows that
\be {\widehat S}_\a'= z_\a -2i {\widehat A}'_\a\ .\label{sprime} \ee
The remaining constraints in $Z$-space, \emph{viz.}
\bea [\widehat S'_\a,\widehat S'_\b]_\star&=&
-2i\e_{\a\b}(1-c_1\widehat \Phi'\star\kappa)\ ,\qquad[\widehat
S'_{\ad},\widehat S'_{\bd}]_\star\ =\  -2i\e_{\ad\bd}(1-c_2\widehat
\Phi'\star\bar\kappa)\ ,\label{z1}\eea\bea [\widehat S'_\a,\widehat
S'_{\bd}]_\star&=&0\ ,\label{z2}\eea \bea \widehat
S'_\a\star\widehat\Phi'+\widehat\Phi'\star\pi(\widehat S'_\a)&=&0\
,\label{z3}\eea\bea \widehat
S'_{\ad}\star\widehat\Phi'+\widehat\Phi'\star\bar\pi(\widehat
S'_{\ad})\ =\ 0\ ,\label{z4}\eea
are then to be solved with an initial condition
 \be
C'(y,\bar y)\ =\ \widehat\Phi'|_{Z=0}\ ,\label{C}
 \ee
and some assumption about the topology of the internal flat
connections
\bea
 \widehat S^{\prime(0)}_{\a}&=&\widehat S'_\a|_{C'=0}\ ,\qquad
 \widehat S^{\prime(0)}_{\ad}\ =\ \widehat S'_{\ad}|_{C'=0}\ .
 \label{physgauge}
\eea

In what follows, we shall restrict the class of solutions further by
assuming that
\be \widehat L\ =\ L(x;y,\yb)\ .\ee
The gauge fields can then be obtained from \eq{Amu}, \eq{Kmu} and
\eq{Leq}, \emph{viz.}
\be e_\m+\o_\m+W_\m\ =\ L^{-1}\partial_\m L-K_\m\ ,\label{gf}\ee
where
\bea K_\m&=&{1\over 4i}\left.L^{-1}\star \left(\o_{\m}{}^{\a\b}
\widehat S'_{\a}\star\widehat S'_{\b}+\bar\o_{\m}{}^{\ad\bd}
\widehat S'_{\ad}\star\widehat S'_{\bd}\right)\star L\right|_{Z=0}\
.\label{Kmu2}\eea
Hence, the gauge fields, including the metric, can be obtained
algebraically without having to solve any differential equations in
spacetime.


\scss{Ordinary Maximally Symmetric Spaces (Type 0)}\label{Sec:symm}


The complex master-field equations are solved by
\be \widehat\Phi\ =\ 0\ ,\qquad \widehat S_\a\ =\ z_\a\ ,\qquad
\widehat S_{\ad}\ =\ \bar z_{\ad}\ ,\qquad \widehat A_\m\ =\
L^{-1}\star
\partial_\m L \ ,\ee
where the gauge function \cite{Bolotin:1999fa}
\bea L(x;y,\yb)&=& {2h\over 1+h} \exp\left[{i\lambda
x^{\a\dot\a}y_\a \bar y_{\dot\a}\over 1+h}\right]\ ,\label{wL} \eea
gives
\be ds^2_{(0)}\ =\ {4 dx^2\over (1-\lambda^2x^2)^2}\
,\label{vacmetric}\ee
which we identify as the metric of the symmetric spaces listed in
Table \ref{Table1} for the different real forms of the model, in
stereographic coordinates with inverse radius $|\lambda|$. This
metric is invariant under the inversion
\bea x^a\rightarrow -x^a/(\l^2 x^2)\ ,\eea
and $H_4$ is covered by a single coordinate chart, while the
remaining symmetric spaces require two charts, related by the
inversion. If we let $\tilde x^a= -x^a/(\l^2 x^2)$, the atlases are
given by
\bea S^4\ \quad(\l^2=-1)&:& \{x^\mu:0\leq -\lambda^2 x^2\leq 1\}\cup
\{\tilde x^\mu:0\leq
-\lambda^2 \tilde x^2\leq 1\}\ ,\label{atl1}\\[5pt]
H^4\quad\quad(\l^2=1)&:& \{x^\mu:0\leq \lambda^2 x^2<1\}\ ,\label{atl2}\\[5pt]
dS_4\quad(\l^2=-1)&:& \{x^\mu:-1< -\lambda^2 x^2\leq 1\}\cup
\{\tilde x^\mu:-1<
-\lambda^2 \tilde x^2\leq 1\}\ ,\label{atl3}\\[5pt]
AdS_4\ \quad(\l^2=1)&:& \{x^\mu:-1 \leq \lambda^2 x^2<1\}\cup
\{\tilde x^\mu:-1\leq
\lambda^2 \tilde x^2<1\}\ ,\label{atl4}\\[5pt]
H_{3,2}\quad(\l^2=-1)&:& \{x^\mu:-1< -\lambda^2 x^2\leq 1\}\cup
\{\tilde x^\mu:-1< -\lambda^2 \tilde x^2\leq 1\}\ ,\label{atl5}\eea
where the overlap between the charts is given by $\{x^\mu:\lambda^2
x^2=-1\}$ in the cases of $S^4$, $dS_4$, $AdS_4$ and $H_{3,2}$, and
the boundary is $\{x^\mu:\lambda^2 x^2=1\}$ in the case of $H_4$ and
$\{x^\mu:\lambda^2 x^2=1\}\cup \{\tilde x^\mu:\lambda^2\tilde
x^2=1\}$ in the cases of $dS_4$, $AdS_4$ and $H_{3,2}$. The
$H_{3,2}$ space can be described as the coset $SO(3,2)/SO(2,2)$.


\scss{$SO(4-p,p)$ Invariant Solutions to the Minimal Model (Type 1)}



\scsss{Internal Master Fields}


A particular class of $SO(4;\Comp)$-invariant solutions is given by
the ansatz
\be \widehat \Phi'\ =\ \nu\ ,\qquad \widehat S'_\alpha\ =\
z_\a~S(u)\ ,\qquad \widehat S'_{\ad}\ =\ \bar z_{\ad}~\bar S(\bar
u)\ee
where
\be u\ =\ y^\a z_\a\ ,\qquad \bar u\ =\ \bar y^{\dot \a} \bar
z_{\dot \a}\ .\ee
The above ansatz solves \eq{z2}-\eq{z4}. There remains to solve
\eq{z1}, which now takes the form
\be [\widehat S^{\prime\alpha},\widehat S'_\a]_\star\ =\ 4i(1-
c_1\nu e^{iu})\ ,\qquad [\widehat S^{\prime\ad},\widehat
S'_{\ad}]_\star\ =\ 4i(1- c_2\nu e^{-i\bar u})\label{SSnu}\ee
Following \cite{Prokushkin:1998bq}, we use the integral
representation
\bea S(u)&=&\int_{-1}^1 ds~ n(s) ~e^{\frac{i}2(1+s)u}\ ,\label{SSnu2}\\
\bar S(\bar u)&=& \int_{-1}^1 ds~ \bar n(s) ~e^{-\frac{i}2(1+s)\bar
u}\ .\label{Su}\eea
which reduces \eq{SSnu} to
\bea (n \circ n)(t)&=&\ \delta(t-1)-\frac{c_1\nu}2 (1-t)\ ,\\
(\bar n \circ \bar n)(t)&=& \delta(t-1)-\frac{c_2\nu}2 (1-t)\ .\eea
with $\circ$ defined by \cite{Prokushkin:1998bq}
 \be
 (f\circ g)(t)=\int_{-1}^1 ds \int_{-1}^1 ds'
 \delta(t-s s')~f(s)~g(s')\ .
 \ee
Even and odd functions, denoted by $f^\pm(t)$, are orthogonal with
respect to the $\circ$ product. Thus, one finds
\bea (n^{+}\circ n^{+})(t)&=&\iota^{+}_0(t)-\frac{c_1\nu}2\ ,\qquad
(n^{-}\circ n^{-})(t)\ =\ \iota^{-}_0(t)+\frac{c_1\nu}2 t\
,\label{mplus}\\(\bar n^{+}\circ \bar
n^{+})(t)&=&\iota^{+}_0(t)-\frac{c_2\nu}2\ ,\qquad (\bar n^{-}\circ
\bar n^{-})(t)\ =\ \iota^{-}_0(t)+\frac{c_2\nu}2 t\
,\label{mminus}\eea
where
\be \iota^{\pm}_0(t)\ =\
\frac12\left[\delta(1-t)\pm\delta(1+t)\right]\ .\ee
One proceeds \cite{Prokushkin:1998bq}, by writing
\bea n^\pm(t)&=& m^\pm(t)+\sum_{k=0}^\infty \l_k p^\pm_k\
,\label{mexp}\eea
where $m^\pm$ are expanded in terms of $\iota^{(\pm)}_0(t)$ and the
functions ($k\geq 1$)
 \bea
 \iota^{\s}_k(t)&=&\left[{\rm sign}(t)\right]^{\frac12(1-\s
 )}~\int_{-1}^1 ds_1 \cdots \int_{-1}^1 ds_k~\delta(t-s_1\cdots
 s_k)\nn\\[5pt]
 &=&\left[{\rm sign}(t)\right]^{\frac12(1-\s)}{\left(\log
 \frac1{t^2}\right)^{k-1}\over (k-1)!}\ ,
 \eea
obeying the algebra ($k,l\geq 0$)
 \be
 \iota^{\s}_k\circ \iota^{\s}_l\ =\ \iota^{\s}_{k+l}\ ,
 \label{ring}
 \ee
and $p^\s_k(t)$ ($k\geq 0$) are the $\circ$-product projectors
\bea p^{\s}_k(t)&=& {(-1)^k\over k!} \d^{(k)}(t)\ ,\qquad \s\ =\
(-1)^k\ ,\label{pk}\eea
obeying
\bea p^\s_k\circ f&=& L_k[f] p^\s_k\ ,\qquad L_k[f]\ =\ \int_{-1}^1
dt~ t^k f(t)\ .\label{proj1}\eea
In particular,
\bea p^{\s}_k\circ p^{\s}_l&=& \delta_{kl}p^{\s}_l\
.\label{proj2}\eea
Substituting the expansion \eq{mexp} into \eq{mplus} and
\eq{mminus}, one finds, in view of \eq{ring}, \eq{proj1} and
\eq{proj2}, manageable algebraic equations. Transforming back one
finds, after some algebra \cite{Sezgin:2005pv},
\bea m(t)&=& \d(1+t)+q(t)\ ,\\[5pt] q(t) &=&-{c_1\nu\over 4}\left({}_1\!
F_1\left[\frac12;2;{c_1\nu\over 2}\log \frac 1{t^2}\right]+t\,{}_1\!
F_1\left[\frac12;2;-{c_1\nu\over 2}\log \frac 1{t^2}\right]\right)\
,\label{mt}\eea
and
\bea \l_k&=&- 2\th_kL_k[m]\ ,\qquad \th_k\in\{0,1\}\
,\label{lambdak}\eea
where
\bea L_k[m]&=&(-1)^k+L_k[q]\ ,\label{Lkm}\\[5pt] L_k[q]&=& -{1+(-1)^k\over
2}\left(1-\sqrt{1-{c_1\nu\over 1+k}}\right)- {1-(-1)^k\over
2}\left(1-\sqrt{1+{c_1\nu\over 2+k}}\right)\ .\eea
The overall signs in $m^\pm$ have been fixed in \eq{mt} by requiring
that
\be S(u)=1\ \ {\rm for}\ \  \nu=0 \ \ {\rm and} \ \ \theta_k=0 \ . \ee
Treating $\bar n$ the same way, one finds
\bea \bar m(t)&=& \d(1+t)+\bar q(t)\ ,\\[5pt] \bar q(t) &=&-{c_2\nu\over 4}\left({}_1\!
F_1\left[\frac12;2;{c_2\nu\over 2}\log \frac 1{t^2}\right]+t\,{}_1\!
F_1\left[\frac12;2;-{c_2\nu\over 2}\log \frac 1{t^2}\right]\right)\
,\label{mt2}\\[5pt]\bar \l_k&=&- 2\bar \th_kL_k[\bar m]\ ,\qquad \bar \th_k\in\{0,1\}\ ,\\[5pt]
L_k[\bar m]&=&(-1)^k+L_k[\bar q]\ ,\\[5pt] L_k[\bar q]&=& -{1+(-1)^k\over
2}\left(1-\sqrt{1-{c_2\nu\over 1+k}}\right)- {1-(-1)^k\over
2}\left(1-\sqrt{1+{c_2\nu\over 2+k}}\right)\ .\eea
Thus, the internal solution is given by
 \bea
 \widehat \Phi'&=&\nu\ ,\label{intsol1}\eea
together with ${\widehat S}'_\a$ and ${\widehat S}'_{\dot\a}$ as
given in \eq{sprime} with
 \bea \widehat A'_\a&=&
 \widehat A^{\prime(reg)}_\a+\widehat A^{\prime(proj)}_\a\ ,\qquad
 \qquad\qquad\qquad\!\! \widehat
 A'_{\ad}\ =\
 \widehat A^{\prime(reg)}_{\ad}+\widehat A^{\prime(proj)}_{\ad}\ ,\label{intconn}\\[5pt]
 \widehat A^{\prime(reg)}_\a&=&\frac{i}2 z_\a \int_{-1}^1 dt~ q(t)\,
 e^{\frac{i}2(1+t)u}\ ,\qquad\,\,\quad \widehat A^{\prime(reg)}_{\ad}
 \ =\ \frac{i}2 \bar z_{\ad} \int_{-1}^1 dt~ \bar q(t)\,
 e^{-\frac{i}2(1+t)\bar u}\ ,\label{part}\\[5pt]\widehat A^{\prime(proj)}_\a&=&
 -iz_\a \sum_{k=0}^\infty \theta_k (-1)^k L_k[m] P_k(u)\ ,\qquad \widehat
 A^{\prime(proj)}_{\ad}\ =\
 -i\bar z_{\ad}\sum_{k=0}^\infty \bar \theta_k (-1)^k L_k[\bar m] \bar P_k(\bar u)\
 ,\qquad \label{hom}\eea
where
\bea P_k(u)&=&\int_{-1}^1 ds ~e^{\ft{i}2 (1-s)u} p_k(s)\ =\ {1\over
k!}\left({-iu\over 2}\right)^ke^{\ft{iu}2}\ ,\label{Pk1}\\[5pt]
\bar P_k(\bar u)&=&\int_{-1}^1 ds ~e^{-\ft{i}2 (1-s)\bar u} p_k(s)\
=\ {1\over k!}\left({i\bar u\over 2}\right)^k e^{-\ft{i\bar
u}2}\label{Pk2}\eea
are projectors in the $\star$-product algebra given by functions of
$u$ and $\bar u$, \emph{viz.}
\bea P_k\star F&=& L_k[f]P_k\ ,\qquad P_k\star P_l\ =\
\delta_{kl}P_k\ ,\\[5pt] \bar P_k\star \bar F&=& L_k[\bar f]\bar P_k\ ,
\qquad \bar P_k\star \bar P_l\ =\
\delta_{kl}\bar P_k\ ,\eea
for $F(u)=\int_{-1}^1 ds e^{\ft{i}2(1-s)u}f(s)$ and $\bar F(\bar
u)=\int_{-1}^1 ds e^{-\ft{i}2(1-s)\bar u}\bar f(s)$ with $L_k[f]$
and $L_k[\bar f]$ given in \eq{proj1}. The projectors also obey
$(u-2ik)\star P_k=0$ and $y^\a\star P_k\star
z_\a=i(k+1)(P_{k-1}+P_{k+1})$ with $P_{-1}\equiv 0$. We note the
opposite signs in front of $s$ in the exponents of \eq{SSnu2},
\eq{Su} and \eq{Pk1}, \eq{Pk2}, resulting in the $(-1)^k$ in the
projector part \eq{hom} of the internal connection, which we can
thus write as
\bea \widehat
A^{\prime(proj)}_\a\!\!\!&=&\!\!\!-iz_\a\sum_{k=0}^\infty\left[\theta_k
P_k-\left(1-\sqrt{1-{c_1\nu\over
1+2k}}\right)\theta_{2k}P_{2k}+\left(1-\sqrt{1+{c_1\nu\over
3+2k}}\right)\theta_{2k+1}P_{2k+1}\right]\,,\qquad\qquad\\[5pt]
\widehat A^{\prime(proj)}_{\ad}\!\!\!&=&\!\!\!-i\bar
z_{\ad}\sum_{k=0}^\infty\left[\bar\theta_k \bar
P_k-\left(1-\sqrt{1-{c_2\nu\over 1+2k}}\right)\bar \theta_{2k}\bar
P_{2k}+\left(1-\sqrt{1+{c_2\nu\over 3+2k}}\right)\bar
\theta_{2k+1}\bar P_{2k+1}\right]\,,\qquad\qquad\eea
which are analytic functions of $\nu$ in a finite region around the
origin. For example, for $c_1=c_2=1$, they are real analytic for
$-3<{\rm Re}\nu<1$, where also the particular solution can be shown
to be real analytic \cite{Sezgin:2005pv}. The reality conditions on
the $\th_k$ and $\bar\th_k$ parameters are as follows:
\bea \mbox{$(4,0)$ and $(2,2)$ signature}&:& \th_k\ ,\quad
\bar\th_k\qquad \mbox{independent}\ ,\\[5pt]
\mbox{$(3,1)$ signature}&:& \th_k\ =\ \bar\th_k\ .\eea

Taking $\nu=0$ there remains only the projector part, leading to the
following ``vacuum'' solutions
\bea \widehat \Phi'&=&0\ ,\eea\bea \widehat
A'_\a&=&-iz_\a\sum_{k=0}^\infty\theta_k {1\over k!}\left({-iu\over
2}\right)^ke^{\ft{iu}2}\ ,\qquad  \widehat A'_{\ad}\ =\ -i\bar
z_{\ad}\sum_{k=0}^\infty\bar\theta_k {1\over k!}\left({i\bar u\over
2}\right)^k e^{-\ft{i\bar u}2}\ .\eea
The $\integ_2\times \integ_2$ symmetry \eq{Z2} acts by
\bea \th_k&\rightarrow& 1-\th_k\ ,\qquad \bar\th_k\ \rightarrow\
1-\bar\th_k\ .\eea
The maximally symmetric spaces discussed in Section 3.2 are
recovered by setting $\th_k=\th$ and $\bar \th_k=\bar\th$ for all
$k$. In Euclidean and Kleinian signatures, $\th$ and $\bar\th$ are
independent, leading to four solutions related by $\integ_2\times
\integ_2$ transformation. In Lorentzian signature, $\th=\bar\th$
leading to two solutions related by $\integ_2$ symmetry.


\scsss{Spacetime Component Fields}


The calculation of the component fields follow the same steps as in
\cite{Sezgin:2005pv}. The spin $s\geq 1$ Weyl tensors vanish, while
the scalar field is given by
\bea\phi(x)&=& \nu h^2(x^2)\ =\ \nu(1-\lambda^2 x^2)\ .\eea
In order to compute the gauge fields, we first need to compute the
quantity $K_\m$ given in \eq{Kmu2}. This calculation is formally the
same as the one spelled out in the case of $\th_k=\bar\th_k=0$ in
\cite{Sezgin:2005pv}, and result is
\bea K_\mu&=&{Q\over 4i}\omega_\mu^{\a\b}v_\a v_\b+{\bar Q\over
4i}\bar\omega_{\mu}^{\ad\bd}\bar v_{\ad}\bar v_{\bd}\ ,\eea
where
\bea Q&=& -{(1-a^2)^2\over 4}\int_{-1}^1ds \int_{-1}^1
ds'{(1+s)(1+s')n(s) n(s')\over (1-ss'a^2)^4}\ ,\label{Q}\\[5pt]
\bar Q&=& -{(1-a^2)^2\over 4}\int_{-1}^1ds \int_{-1}^1
ds'{(1+s)(1+s')\bar n(s) \bar n(s')\over (1-ss'a^2)^4}\
.\label{barQ}\eea
and
\bea v_\a&=&(1+a^2)y_\a+2(a\bar y)_\a\ ,\qquad \bar v_{\ad}\ =\
(1+a^2)\bar y_{\ad}+2(\bar a y)_{\ad}\ ,\eea
with ${\bar a}_{\ad \a}= a_{\a\ad}$ defined in \eq{aad}. We can
simplify $Q$ using $n(t)=\delta(1+t)+q(t)+\sum_k\lambda_k p_k(t)$,
with $p_k(t)$ given by \eq{pk} and $\lambda_k$ by \eq{lambdak} and
\eq{Lkm}. After some algebra we find
\bea Q(\nu;\{\th_k\})&=&Q^{(reg)}(\nu)+Q^{(proj)}(\nu;\{\th_k\})\ ,\\[5pt]
Q^{(reg)}&=& -{(1-a^2)^2\over 4}\int_{-1}^1ds \int_{-1}^1
ds'{(1+s)(1+s')q(s) q(s')\over (1-ss'a^2)^4}\ ,\\[5pt]
Q^{(proj)} &=& (1-a^2)^2\sum_{k=0}^\infty {4_k a^{2k}\over
k!}(\th_k-\th_{k+1})^2\left((-1)^k+L_k(q)\right)\left((-1)^{k+1}+L_{k+1}(q)\right)\
,\eea
where we note that $Q$ depends on $\th_k$ only via
$\th_k-\th_{k+1}$. The same expression with $q\rightarrow\bar q$ and
$\th_k\rightarrow \bar\th_k$ holds for $\bar Q$. The regular part,
which was computed in \cite{Sezgin:2005pv}, is given by
\bea Q^{(reg)}&=& Q^{(reg)}_++Q^{(reg)}_-\ ,\\[5pt]
Q^{(reg)}_+&=&-{(1-a^2)^2\over 4}\sum_{p=0}^\infty {-4\choose
2p}a^{4p}\left(\sqrt{1-{c_1\nu\over 2p+1}}-\sqrt{1+{c_1\nu\over
2p+3}}\right)^2\label{Qpl}\\[5pt]
Q^{(reg)}_-&=&{(1-a^2)^2\over 4}\sum_{p=0}^\infty {-4\choose
2p+1}a^{4p+2}\left(\sqrt{1-{c_1\nu\over 2p+3}}-\sqrt{1+{c_1\nu\over
2p+3}}\right)^2\ ,\label{Omi}\eea
while a similar expression, obtained by replacing $c_1\rightarrow
c_2$, holds for $\bar Q$.

Since $K_\mu$ is bilinear in the $y_\a$ and $\yb_{\ad}$ oscillators,
it immediately follows that all higher spin fields vanish. Moreover,
after some algebra, we find that the vierbein and $\mso(4;\Comp)$
connection are given by
\bea e^a&=& f_1(x^2) dx^a+f_2(x^2) x^a dx^b x_b\ ,\\[5pt]
\omega_{\a\b}&=& f(x^2)\omega^{(0)}_{\a\b}\ ,\qquad
\bar\omega_{\ad\bd}\ =\ \bar f(x^2) \bar\omega^{(0)}_{\ad\bd}\ ,\eea
where
\bea f&=& {1+(1-a^2)^2\bar Q\over
\left[1+(1+a^2)^2Q\right]\left[1+(1+a^2)^2\bar
Q\right]-16a^4 Q\bar Q}\ ,\\[5pt] \bar f&=& {1+(1-a^2)^2 Q\over \left[1+(1+a^2)^2Q\right]\left[1+(1+a^2)^2\bar
Q\right]-16a^4 Q\bar Q}\ ,\eea
and
\bea f_1+\lambda^2 x^2 f_2&=& {2\over h^2}\ ,\qquad f_2\ =\
{2(1+a^2)^4\over (1-a^2)^2}(fQ+\bar f\bar Q)\ .\label{f12}\eea
By a change of coordinates, the metric can be written locally, in a
given coordinate chart, as a foliation
\bea ds^2&=& \e d\tau^2+R^2 d\O_3^2\ ,\qquad R^2(\tau)\ =\ \eta^2
|\sinh^2(\sqrt{\e}\tau)|\ ,\eea
where $x^2=\e \tan^2\ft{\tau}2$ with $\e=\pm 1$, and $d\Omega_3^2$
is a three-dimensional metric of constant curvature with suitable
signature, and \cite{Sezgin:2005hf}
\bea \eta&=&{f_1 h^2\over 2}\ .\label{etafactor}\eea
One has the following simplifications in specific models:
\bea \mbox{Type A model:}&& Q\ =\ \bar Q\ ,\qquad \eta\ =\
{1+(1-a^2)^2Q\over 1+(1+6a^2+a^4)Q}\ ,\\[5pt]
\mbox{Chiral model:}&& \bar Q\ =\ 0\ ,\qquad \eta\ =\
{1+(1-a^2)^2Q\over 1+(1+a^2)^2Q}\ .\eea
The metric may have conical singularities, namely zeroes
$R(\tau_0)=0$ for which $\partial_\tau R|_{\tau_0}\neq 1$ (we note
that $\eta|_{\tau=0}=1$, so that $\tau=0$ is not a conical
singularity). The scale factor depend heavily on $\nu$ as well as
the choice of the infinitely many discrete parameters $\th_k$ and
$\bar\th_k$. This makes the analysis unyielding, and we shall
therefore limit ourselves to the case of vanishing discrete
parameters and $\nu\ll 1$. In Lorentzian signature, the resulting
analysis was performed in \cite{Sezgin:2005hf}, and it generalizes
straightforwardly to Euclidean and Kleinian signatures. To this end,
one expands $Q^{(reg)}$ in $\nu$ around $\nu=0$, and finds
\bea Q^{(reg)}&=&{c_1\nu\over 6}\log(1+a^2)+{\cal O}(\nu^2)\ .\eea
Focusing on a single chart, as listed in \eq{atl1}-\eq{atl5}, since
$a^2$ is then bounded from below by $(1-\sqrt{2})(1+\sqrt{2})^{-1}$,
we see that, if $|\nu|\ll 1$, then $|Q|\ll 1$, and consequently the
factor $\eta$ defined in \eq{etafactor} remains finite. Thus, for
small enough $\nu$, there are no conical singularities within the
coordinate charts. However, they may appear for some finite critical
$\nu$.

While the $Q$ functions are highly complicated for $\nu\neq 0$, they
simplify drastically at $\nu=0$, where we find
\bea Q&=&-(1-a^2)^2\sum_{k=0}^\infty {4_k a^{2k}\over
k!}(\th_k-\th_{k+1})^2\ .\eea
An analogous expression can be found for $\bar Q$. Setting
$(\th_k-\th_{k+1})^2=1$, yields
\bea Q&=&-{1\over (1-a^2)^2}\ .\eea
If $Q=\bar Q=-(1-a^2)^{-2}$, which is necessarily the case in the
Lorentzian models, then the equation system for $\o_{\a\b}$ and
$\bar\o_{\ad\bd}$ becomes degenerate, and one finds
\bea \o_{\a\b}&=& -{(1-a^2)^2\over 8a^2} \omega^{(0)}_{\a\b}\ =\
{(\s^{ab})_{\a\b}dx_a x_b\over 2 x^2}\ ,\\[5pt]
\bar\o_{\ad\bd}&=& -{(1-a^2)^2\over 8a^2} \bar\omega^{(0)}_{\ad\bd}\
=\ {(\bar\s^{ab})_{\ad\bd}dx_a x_b\over 2 x^2}\ ,\eea
leading to the degenerate vierbein
\bea e_{\a\ad}&=&-{\l x_{\a\ad} x^a dx_a\over x^2h^2}\ ,\eea
and metric
\bea ds^2&=&{4(x^a dx_a)^2\over\l^2 x^2 h^2}\ .   \eea
%


\scss{Solutions of Non-minimal Model (Type 2)}\label{sec:van0form}



\scsss{Internal Master Fields}


The non-minimal model admits the following solutions
\bea \widehat \Phi'&=&0\ ,\qquad \widehat S'_\a\ =\ z_\a\star
\C(y,\bar y)\ ,\qquad\widehat S'_{\ad}\ =\ \bar z_{\ad}\star \bar
\C(y,\bar y)\ ,\label{type2}\eea
provided that
\bea \C\star\C&=&\bar\C\star\bar\C\ =\ 1\ ,\qquad [\C,\bar
\C]_\star\ =\ 0\ ,\qquad \pi\bar\pi(\C)\ =\ \C\ ,\qquad
\pi\bar\pi(\bar\C)\ =\ \bar\C\ .\label{gprop}\eea
The elements $\C$ and $\bar\C$ can be written as
\bea \C&=&1-2P\ ,\qquad \bar \C\ =\ 1-2\bar P\ ,\eea
where $P(y,\bar y)$ and $\bar P(y,\bar y)$ are projectors obeying
\bea P\star P&=&P\ ,\qquad \bar P\star\bar P\ =\ \bar P\ ,\quad
[P,\bar P]_\star\ =\ 0\ ,\qquad \pi\bar\pi(P)\ =\ P\ ,\qquad
\pi\bar\pi(\bar P)\ =\ \bar P\ .\qquad\eea
A set of such projectors is described in Appendix \ref{AppProj},
where we also explain why the projectors can be subject to the
$\tau$-conditions of the non-minimal model, given in \eq{nonminmod},
but not those of the minimal model, given in \eq{minmod}, unless one
develops some further formalism for handling certain divergent
$\star$-products.


\scsss{Spacetime Component Fields}


Turning to the computation of the space components of the master
fields, since $z_\a$ star-commutes with $L$, it immediately follows
from \eq{gf}, \eq{Kmu} and \eq{type2} that
\bea K_\mu&=&0\ .\eea
From \eq{gf} this in turn implies that all HS gauge fields and the
spin-1 gauge field vanish, while the metric is that of maximally
symmetric spacetime. To that extent, the Type 1 solution looks like
the Type 0 solution, but it does differ in an important way, namely,
here the internal connection, \emph{i.e.} the spinor component
${\widehat A}_\a$ of the master $1$-form, is non-vanishing. Indeed,
\eq{type2}, \eq{gprop} and \eq{sprime} give the result
\bea{\widehat A}_\a&=& -iz_\alpha \star V(x;y,\yb)\ ,\qquad \widehat
A_{\ad}\ =\ -i\bar z_{\ad} \star \bar V(x;y,\yb)\
,\label{spinorform}\eea
where the quantities $V$ and $\bar V$, which shall be frequently
encountered in what follows, are defined by
\bea V&=& L^{-1}\star P\star L\ ,\qquad \bar V\ =\  L^{-1}\star \bar
P\star L \ .\eea
Their explicit evaluation is given in Appendix \ref{App2}, with the
result \eq{generalV}.

Whilst the internal connection does not turn on any spacetime
component fields, it does, however, affect the interactions as it
does not obey the physical gauge condition normally used in the
weak-field expansion \cite{Sezgin:2002ru}, namely that the internal
connection should vanish when the zero-form vanishes. In this sense,
the internal connection may be viewed as a non-trivial flat
connection in the non-commutative space.


\scss{Solutions of Non-minimal Chiral Model (Type 3)}



\scsss{Internal Master Fields}


In the case of the non-minimal chiral model, defined in Section 2.3,
it is possible to use projectors $P(y,\bar y)$ to build solutions
with non-vanishing Weyl zero-form and higher spin fields. They are
\bea\widehat \Phi'&=& (1-P)\star\kappa\ ,\qquad \widehat S'_\a\ =\
z_\a\star P\ ,\qquad \widehat S'_{\ad}\ =\ \bar z_{\ad}\star \bar\C\
,\eea
where
\bea P\star P&=&P\ ,\qquad \bar\C\star\bar\C\ =\ 1\ ,\qquad
[P,\bar\C]_\star\ =\ 0\ ,\qquad \pi\bar\pi(P)\ =\ P\ ,\qquad
\pi\bar\pi(\bar\C)\ =\ \bar\C\ .\qquad\eea
These elements of the $\star$-product algebra can be constructed as
in Section \ref{sec:van0form} and Appendix \ref{AppProj}.

For the purpose of exhibiting explicitly the spacetime component
fields, we shall choose to work with the simplest possible
projectors, namely
\bea P_+(y)&=& 2e^{-2\e uv}\ =\ 2e^{\e yby}\ ,\label{Pplus}\\[5pt]
P_-(\yb)&=& 2e^{-2\e \bar u\bar v}\ =\ 2e^{\e \bar y\bar b \bar y}\
,\label{Pminus}\eea
where $\e=\pm 1$, and $u$, $v$, $\bar u$, $\bar v$, $b_{\a\b}$ and
$\bar b_{\ad\bd}$ are defined in Appendices \ref{AppDef} and
\ref{AppProj}.


\scsss{Spacetime Component Fields}


The master gauge field and zero-form is given by
\bea e_\mu+\omega_\mu+W_\mu&=&
e^{(0)}_\mu+\omega^{(0)}_\mu+{\o_\mu^{\a\b}\over 4i} {\partial^2 V
\over\partial y^\a\partial y^\b}\ ,\label{Type3}\eea
and
\bea\Phi&=&\left[L^{-1}\star(1-P)\star\kappa\star
\pi(L)\right]|_{Z=0}\ =\ 1-V|_{y_\a=0}\ ,\eea
where $K_\mu$ is defined by \eq{Kmu}; we have used \eq{kappa}; and
$V$ is given by \eq{generalV}. Remarkably, since there is no
$y$-dependence in the Weyl zero-form $\Phi$, it is covariantly
constant in the sense that $\Phi_{\a(m)\ad(n)}$ vanishes unless
$m=0$. Moreover, using \eq{generalV}, it is straightforward to
compute the constant value of the physical scalar field, with the
result
\bea \phi(x)&=& 1-4\sum_{n_1,n_2\in
\integ+\ft12}(-1)^{n_1+n_2-\ft{\e_1+\e_2}2}\th_{n_1,n_2}\
.\label{VEV}\eea
Summing over all $n_2$, and using \eq{generating} with $x=0$,
\emph{i.e.} $\sum_{k=0}^\infty (-1)^k=\ft12$, one finds that for the
reduced projector \eq{reduced}, the scalar field is given by
\bea \phi(x)&=& 1-2\sum_{n\in \integ+\ft12}(-1)^{n-\ft{\e}2}\th_{n}\
.\eea
Finally setting all $\theta$-parameters equal to $1$, one ends up
with $P=1$, \emph{i.e.} in the Type 0 case, where indeed
$\phi(x)=0$.

In the special cases of \eq{Pplus} and \eq{Pminus}, one finds
\bea V_+ &=& L^{-1}\star P_+\star L\ =\ 2 \exp \left(-\e{ [2\yb \bar
a -(1+a^2)y]\,b\,[2a\yb
+(1+a^2)y]\over (1-a^2)^2}\right)\ ,\label{vnonmin}\\[5pt]
V_-&=&L^{-1}\star P_-\star L\ =\ 2 \exp \left(-\e{ [2y a
-(1+a^2)\yb]\,\bar b\,[2\bar ay +(1+a^2)\yb]\over (1-a^2)^2}\right)
\label{vminus}\eea
where $a_{\a\ad}$ and $b_{\a\b}$ are defined in Appendix
\ref{AppDef}. The physical scalar is now given in both cases by
\bea\phi(x)&=&-1\ ,\eea
and the self-dual Weyl tensors in both cases by ($s=1,2,3,....$)
\bea \Phi_{\a(2s)}&=& 0\ ,\eea
while the anti-self-dual Weyl tensors take the form
\bea \Phi^+_{\ad_1\cdots\ad_{2s}}&=& -2^{2s+1}(2s-1)!!\left( {h^2-1
\over \e h^2}\right)^s\, U_{(\ad_1}\cdots U_{\ad_s}\,
V_{\ad_{s+1}}\cdots V_{\ad_{2s})}\ ,\label{phi+}\\[5pt]
\Phi^-_{\ad_1\cdots\ad_{2s}}&=&-2^{2s+1}(2s-1)!!\left( {1\over \e
h^2}\right)^s\, \bar\l_{(\ad_1}\cdots \bar\l_{\ad_s}\,
\bar\mu_{\ad_{s+1}}\cdots \bar\mu_{\ad_{2s})}\ ,\label{phi-}\eea
with spinors $(U, V)$ defined in \eq{UV}.

In the case of $\l^2=1$ in Euclidean signature, we only need to use
one coordinate chart, in which $0\leq h^2\leq 1$. The Weyl tensors
blow up in the limit $h^2\rightarrow 0$, preventing the solution
from approaching $H_4$ in this limit. In this sense the above
solution is a non-perturbative solution without weak-field limit in
any region of spacetime. Indeed, in the perturbative weak-field
expansion around the $H_4$ solution, the scalar field has
non-vanishing mass, preventing the linearized scalar field from
being a non-vanishing constant.

In the case of $\l^2=-1$ in Euclidean signature, the base manifold
consists of two charts, covered by the coordinates in \eq{atl1}.
Thus, in each chart we have $1\leq h^2<2$, and so the local
representatives \eq{phi+} and \eq{phi-} of the Weyl tensors are
well-defined throughout the base manifold.

Finally, in the case of $\l^2=-1$ in Kleinian signature, one also
needs two charts (since we are working with stereographic
coordinates), with $0\leq h^2\leq 2$, and hence the Weyl tensors
blow up in the limit $h^2\rightarrow 0$ preventing the solution from
approaching $H_{3,2}$ in this limit.

From the Weyl tensors, which are not in themselves HS gauge
invariant quantities, one can construct an infinite set of invariant
(and thus closed) zero-forms \cite{Sezgin:2005pv}, namely
\bea {\cal C}^-_{2p}&=& \int {d^4y d^4z\over (2\pi)^4} [(\widehat
\Phi\star\pi(\widehat\Phi)]^{\star p}\star\kappa\bar\kappa\ .\eea
Remarkably, on our solution they all assume the same value, given by
the constant value of the scalar field, \emph{viz.}
\bea {\cal C}^-_{2p}&=&(1-V)^{\star 2p}|_{y=\bar y=0}\ =\
1-4\sum_{n_1,n_2\in
\integ+\ft12}(-1)^{n_1+n_2-\ft{\e_1+\e_2}2}\th_{n_1,n_2}\ .\eea

The calculation of the metric in the two models  proceeds in a
parallel fashion as follows:

%
{\bf The $P_+$ Solution:}
%
%

From \eq{Type3} and \eq{vnonmin} a straightforward computation
yields the result
\bea e_{\mu\ad\a} &=& e_{\mu\ad\a}^{(0)} +12(1+h)h^{-4}\,
b_{(\a\b}(ba)_{\gamma)\ad}\,\omega_\mu^{\beta\gamma}\
,\label{nm1}\w2
\omega_{\mu\a\b} &=& \omega_{\mu\a\b}^{(0)} +12
h^{-4}b_{(\a\b}b_{\gamma\delta)}\,\omega_\mu^{\gamma\delta}\
,\label{nm2}\w2
{\bar\omega}_{\mu\ad\bd}&=& {\bar\omega}_{\mu\ad\bd}^{(0)}
+4(1+h)^2h^{-4}\left[ -(\bar aba)_{\ad\bd} b_{\gamma\delta} +2 (\bar
ab)_{\ad\gamma} (\bar ab)_{\bd\delta}
\right]\omega_\mu^{\gamma\delta}\ .\label{nm3} \eea
First we solve for the spin connection from \eq{nm2} by inverting
the hyper-matrix that multiplies $\omega^{(0)}$, obtaining the
result
\be \omega_{\mu\a\b}= g_1\left[ \omega_{\mu\a\b}^{(0)} -8g
(b\omega_\mu^{(0)}b)_{\a\b}\right] + g_2 b_{\a\b} b^{\gamma\delta}
\omega_{\mu\gamma\delta}^{(0)}\ ,  \label{nm4}\ee
where
\be g_1={1\over 1-4g^2}\ ,\qquad g_2={4g\over (1-2g)(1-4g)}\ ,\qquad
g=h^{-4}\ . \ee
Substituting this result in \eq{nm1} then gives the vierbein
\be e_\mu^a= {-2\over h^2(1+2g)}\left[ g_3\delta_\mu^a +g_4
\lambda^2 x_\mu x^a + g_5\lambda^2 (Jx)_\mu (Jx)^a\right]\ , \ee
where
\be g_3=1+2h^{-2}\ , \qquad g_4 = 2g\, \qquad g_5= {6g\over 1-4g}\ ,
\label{g345}\ee
and the spin connections are given in \eq{nm4} and \eq{nm3}. Thus,
the metric $g_{\mu\nu}=e_\mu^a e_\nu^b\,\eta_{ab}$ takes the form
\be g_{\mu\nu}= {4\over h^4(1+2g)^2}\,\left[ g_3^2 \eta_{\mu\nu} +
g_4(\lambda^2 x^2 g_4+2g_3) x_\mu x_\nu +g_5(\lambda^2x^2
g_5+2g_3)(Jx)_\mu (Jx)_\nu\right]\ .\ee
The vierbein thus has potential singularities at $h^2=0$ and
$h^2=2$. The limit $h^2\rightarrow 0$ is a boundary in the case of
$\l^2=1$ in Euclidean signature and $\l^2=-1$ in Kleinian signature.
At these boundaries $e_\mu{}^a\sim h^{-2}x_\mu x^a$, \emph{i.e.} a
scale factor times a degenerate vierbein. In the limit
$h^2\rightarrow 2$ one approaches the boundary of a coordinate chart
in the case of $\l^2=1$ in Euclidean signature and $\l^2=-1$ in
Kleinian signature. Also in this limit, the vierbein becomes
degenerate, \emph{viz.} $e_\mu{}^a\sim h^{-2}(Jx)_\mu (Jx)^a$.
{\bf The $P_-$ Solution:}

A parallel computation that uses \eq{Type3} and \eq{vminus} yields
the result
\bea e_{\mu\ad\a} &=& e_{\mu\ad\a}^{(0)} +12\lambda^2 x^2 (1+h)
h^{-4}\,  {\tilde b}_{(\a\b}({\tilde
ba})_{\gamma)\ad}\,\omega_\mu^{\beta\gamma}\ ,\label{nnm1}\w2
\omega_{\mu\a\b} &=& \omega_{\mu\a\b}^{(0)} +12(\lambda^2x^2)^2
h^{-4}{\tilde b}_{(\a\b}{\tilde
b}_{\gamma\delta)}\,\omega_\mu^{\gamma\delta}\ ,\label{nnm2}\w2
{\bar\omega}_{\mu\ad\bd}&=& {\bar\omega}_{\mu\ad\bd}^{(0)}
+4(1+h)^2h^{-4}\left[ -(\bar a{\tilde b}a)_{\ad\bd} {\tilde
b}_{\gamma\delta} +2 ({\tilde b}a)_{\gamma\ad} ({\tilde
b}a)_{\delta\bd} \right]\omega_\mu^{\gamma\delta}\ ,\label{nnm3}
\eea
where ${\tilde b}_{\a\b}$ is defined in \eq{btilde}. As before,
solving for the spin connection from \eq{nnm2} by inverting the
hyper-matrix that multiplies $\omega^{(0)}$, we obtain
\be \omega_{\mu\a\b}= {\tilde g}_1\left[ \omega_{\mu\a\b}^{(0)}
-8{\tilde g} ({\tilde b}\omega_\mu^{(0)}{\tilde b})_{\a\b}\right] +
{\tilde g}_2 {\tilde b}_{\a\b} {\tilde b}^{\gamma\delta}
\omega_{\mu\gamma\delta}^{(0)}\ , \label{nnm4}\ee
where
\be {\tilde g}_1={1\over 1-4{\tilde g}}\ ,\qquad {\tilde
g}_2={4{\tilde g}\over (1-2{\tilde g})(1-4{\tilde g})}\ ,\qquad
{\tilde g}= (\lambda^2x^2)^2 h^{-4}\ . \ee
Substituting this result in \eq{nnm1} then gives the vierbein
\be e_\mu^a= {-2\over h^2\left(1+2{\tilde g} \right)}\left[
\delta_\mu^a +{\tilde g}_4 \lambda^2 x_\mu x^a + {\tilde
g}_5\lambda^2 ({\tilde J}x)_\mu ({\tilde J}x)^a\right]\ , \ee
where ${\tilde J}_{ab}$ is defined in \eq{jtilde}
\be {\tilde g}_4 = 2\lambda^2 x^2h^{-4}\, \qquad {\tilde g}_5=
{6\lambda^2 x^2h^{-4}\over 1-4{\tilde g}}\ , \label{tildeg345}\ee
and the spin connections are given in \eq{nnm4} and \eq{nnm3}. Thus,
the metric $g_{\mu\nu}=e_\mu^a e_\nu^b\,\eta_{ab}$ takes the form
\be g_{\mu\nu}= {4\over h^4\left[1+2{\tilde g}\right]^2}\,\left[
\eta_{\mu\nu} + {\tilde g}_4(\lambda^2 x^2 {\tilde g}_4+2) x_\mu
x_\nu +{\tilde g}_5(\lambda^2x^2 {\tilde g}_5+2)({\tilde J}x)_\mu
({\tilde J}x)_\nu\right]\ .\ee
The vierbein has potential singularities at $h^2=0$, $h^2=2$ and
$h^2=\ft23$. The singularities at $h^2=0$ and $h^2=2$ are related to
degenerate vierbeins exactly as for the $P^+$ solution. The
singularity at $h^2=\ft23$, which arises in the case of $\l^2=1$ in
Euclidean and Kleinian signature, also gives a degenerate vierbein.
This is an intriguing situation since the degeneration occurs inside
the coordinate charts.


\scs{Conclusions}


Starting from HS gauge theories in four dimensions based on infinite
dimensional extensions of $SO(5;\Comp)$, we have determined their
real forms in spacetimes with Euclidean $(4,0)$ and Kleinian $(2,2)$
signature, in addition to the usual Lorentzian $(3,1)$ signature. We
have then found three new types of solutions in addition to the
maximally symmetric ones. Type 1 solutions, which are invariant
under an infinite dimensional extension of $SO(4-p,p)$, give us a
nontrivial deformation of the maximally symmetric solutions, and
depend on a continuous real parameter as well as infinite set of
discrete parameters. Interestingly, a particular choice of the
discrete parameters, in the limit of vanishing continuous parameter,
gives rise to a degenerate, indeed rank one, metric. Given that
degenerate metrics are known to play an important role in topology
change in quantum gravity \cite{Horowitz:1990qb}, it is remarkable
that such metrics emerge naturally in HS gauge theory.

Type 2 solutions, which provide another kind of deformation of the
maximally symmetric solutions, have a non-vanishing spinorial master
one-form. Type 3 solutions are particularly remarkable because all
the higher spin fields are non-vanishing, and the corresponding Weyl
tensors furnish a higher spin generalization of Type D gravitational
instantons. It would be interesting to apply the framework we have
used in this paper to finding pp-wave, black hole and domain wall
solutions with non-vanishing HS fields.

We stress that our models in Euclidean and Kleinian signatures are
formulated using the 4D spinor-oscillator formulation. It would be
interesting to compare these models to the vector-oscillator
formulation \cite{Vasiliev:2003ev}. The latter exists in any
dimension and signature, and relies on the gauging of an internal
$Sp(2)$ gauge symmetry. At the full level, the vector-oscillator
master field equations, in any dimension and signature, are
formulated using a single $Sp(2)$-doublet $Z$-oscillator, leaving,
apparently, no room for parity violating interactions. The precise
relation between the spinor and vector-oscillator formulations in
D=4 therefore deserve further study.

In the context of supersymmetric field theories, including
supergravity, the non-Lorentzian signature typically presents
obstacle since the spinor properties are sensitive to the spacetime
signature. Here, however, we have considered bosonic HS gauge
theories in which the spinor oscillators play an auxiliary role, and
we have formulated the non-Lorentzian signature theories with
suitable definition of the spinors without having to face such
obstacles. Remarkably, non-supersymmetric 4D theories in Kleinian
signature describing self-dual gravity arise in worldsheet $N=2$
supersymmetric string theories, known as $N=2$ strings. For reasons
mentioned in the introduction, it is an interesting open problem to
find a niche for Kleinian HS gauge theory in a variant of an $N=2$
string.

There are several other open problems that deserve investigation. To
begin with, we have not determined the symmetries of Type 2 and Type
3 solutions. While it may be useful in its own right to determine
whether our Type 3 solutions support a complex, possibly K\"ahler,
structure up to a conformal scaling, such results may be limited in
shedding light to the geometry associated with infinitely many gauge
fields present in HS gauge theory. The correct interpretation of the
singularities or degeneracies in the metrics we have found also
require sufficient knowledge of the HS geometry. Furthermore, a
proper formulation of the HS geometry would also provide a framework
for constructing invariants that could distinguish the gauge
inequivalent classes of exact solutions.

It would also be interesting to study the fluctuations about our
exact solutions, and explore their potential application in quantum
gravity and cosmology. Similarities between the frameworks for
studying instanton and soliton solutions of the noncommutative field
theories (see, for example, \cite{Schaposnik:2003vr}), and in
particular open string field theory, are also worth investigating.

\bigskip\bigskip

{\bf Acknowledgements}

We are grateful to J. Engquist and A. Sagnotti for discussions. P.S.
thanks the Physics Department at Texas A\& M University, and E.S.
thanks the Scuola Normale Superiore, where part of this work was
done, for hospitality. The research of C.I. was supported in part by
INFN; by the MIUR-PRIN contract 2003-023852; by the EU contracts
MRTN-CT-2004-503369 and MRTN-CT-2004-512194; by the INTAS contract
03-51-6346; and by the NATO grant PST.CLG.978785. The research of
E.S. was supported in part by NSF Grant PHY-0555575. The research of
P.S. was supported in part by a visiting professorship issued by
Scuola Normale Superiore; by INFN; by the MIUR-PRIN contract
2003-023852; by the EU contracts MRTN-CT-2004-503369 and
MRTN-CT-2004-512194; by the INTAS contract 03-51-6346; and by the
NATO grant PST.CLG.978785.

\newpage


\begin{appendix}


\scs{General Conventions and Notation}\label{App1}


We use the conventions of \cite{Sezgin:1998gg} in which the real
form of the $SO(5;\Comp)$ generators obey
\be [M_{AB},M_{CD}]\ =\ i\y_{BC}M_{AD}+\mbox{$3$ more}\ ,\qquad
(M_{AB})^\dagger\ =\ \sigma(M_{AB})\ ,\label{sogena}\ee
where $\eta_{AB}=(\eta_{ab};-\lambda^2)$. The commutation relations
above decompose as
\be [M_{ab},M_{cd}]_\star\ =\ 4i\y_{[c|[b}M_{a]|d]}\ ,\qquad
[M_{ab},P_c]_\star\ =\ 2i\y_{c[b}P_{a]}\ ,\qquad [P_a,P_b]_\star\ =\
i\lambda^2 M_{ab}\ .\label{sogenb}\ee
The corresponding oscillator realization is taken to be
 \be
 M_{ab}\ =\ -\frac18 \left[~ (\s_{ab})^{\a\b}y_\a y_\b+
 (\sb_{ab})^{\ad\bd}\bar y_{\ad}\yb_{\bd}~\right]\ ,\qquad P_{a}\ =\
 \frac{\l}4 (\s_a)^{\a\bd}y_\a \yb_{\bd}\ .\label{mab}
 \ee
Our spinor conventions are
\be \e^{\a\b}\e_{\c\d} \ = \ 2 \d^{\a\b}_{\c\d} \ , \qquad
\e^{\a\b}\e_{\a\c} \ = \ \d^\b_\c \ ,\ee
and
\be (\e_{\a\b})^\dagger \ = \ \left\{ \begin{array}{ll}
\e^{\a\b}& \mbox{for $SU(2)$} \\[5pt]
\e_{\ad\bd}& \mbox{for $SL(2,\Comp)$}  \\[5pt]
\e_{\a\b} & \mbox{for $Sp(2)$}  \end{array} \right. \ee
Oscillator indices are raised and lowered according to the following
conventions, $A^\a=\epsilon^{\a\b}A_\b$, $A_\a=A^\b\epsilon_{\b\a}$.
The reality conditions on oscillators have been summarized in
(\ref{su2}), (\ref{sl2}) and (\ref{sp2}). The van der Waerden
symbols obey
 \bea
  (\s^{a})_{\a}{}^{\ad}(\sb^{b})_{\ad}{}^{\b}&=& \y^{ab}\d_{\a}^{\b}\
 +\ (\s^{ab})_{\a}{}^{\b} \ ,\qquad
 (\sb^{a})_{\ad}{}^{\a}(\s^{b})_{\a}{}^{\bd}\ =\ \y^{ab}\d^{\bd}_{\ad}\
 +\ (\sb^{ab})_{\ad}{}^{\bd} \ ,\label{so4a}\w2
 \ft12 \e_{abcd}(\s^{cd})_{\a\b}&=& \e (\s_{ab})_{\a\b}\ ,\qquad \ft12
 \e_{abcd}(\sb^{cd})_{\ad\bd}\ =\  -\e (\sb_{ab})_{\ad\bd}\ ,\label{so4b}
\eea
where $\e=\sqrt{\det \eta_{ab}}$, and the following reality
conditions
\bea ((\s^a)_{\a\bd})^\dagger\ =\ \left\{ \begin{array}{ll}
-(\sb^a)^{\bd\a} \ = \ -(\s^a)^{\a\bd}&\mbox{for $SU(2)$} \\[5pt]
(\sb^a)_{\ad\b} \ = \ (\s^a)_{\b\ad} &\mbox{for $SL(2,\Comp)$}  \\[5pt]
(\sb^a)_{\bd\a} \ = \ (\s^a)_{\a\bd} &\mbox{for $Sp(2)$} \end{array}
\right.\eea
and
\bea ((\s^{ab})_{\a\b})^\dagger\ =\ \left\{ \begin{array}{ll}
(\s^{ab})^{\a\b} &\mbox{for $SU(2)$} \\[5pt]
(\sb^{ab})_{\ad\bd}  &\mbox{for $SL(2,\Comp)$}  \\[5pt]
(\s^{ab})_{\a\b} &\mbox{for $Sp(2)$} \end{array} \right. \ .\eea
Convenient representations are:
\bea  SU(2)&: &  \qquad \s^a=(i,\s^i) \ ,\qquad \sb^a=(-i,\s^i)  \
,\qquad
\e=i\s^2 \ ; \\[5pt]
 SL(2,\Comp)&: & \qquad \s^a=(-i\s^2,-i\s^i\s^2) \ ,\qquad
\sb^a=(-i\s^2,i\s^2\s^i) \ , \qquad\e=i\s^2 \ ; \\[5pt]
Sp(2)& : & \qquad \s^a=(1,\ts^i) \ , \qquad\sb^a=(-1,\ts^i) \
,\qquad \e=i\s^2 \ ,\eea
where in the last case $\ts^i = (\s^1,i\s^2,\s^3)$. Combining
(\ref{em}) with (\ref{mab}), the real form of the
$\mso(5;\Comp)$-valued connection $\O$ can be expressed as
 \be
  \O\ =\ \frac1{4i}
 dx^\mu\left[\omega_\mu^{\a\b}~y_\a y_\b
 +\bar{\omega}_\mu{}^{\dot\a\dot\b}~{\bar y}_{\dot\a}{\bar y}_{\dot\b}
 + 2 e_\mu^{\a\dot\b}~y_\a {\bar y}_{\dot\b}\right]\
 ,\label{Omega}
 \ee
where
 \be
 \o^{\a\b}\ =\ -\ft14(\s_{ab})^{\a\b}~\o^{ab}\ ,
 \quad
 \ob^{\dot\a\dot\b}\ =\ -\ft14({\sb}_{ab})^{\dot\a\dot\b}~\o^{ab}\ ,
 \quad
 e^{\a\dot\a}\ =\ \ft{\lambda}2(\s_{a})^{\a \dot\a}~e^{a}\ .
 \label{convert}
 \ee
Likewise, for the curvature ${\cal R}=d\O+\O\wedge\star \O$ one
finds
 \bea
 {\cal R}_{\a\b}&=& d\o_{\a\b} +\o_{\a\c}\wedge\o_{\b}{}^{\c}+
 e_{\a\dd}\wedge e_{\b}{}^{\dd}\ ,
 \label{rab}\w2
 \bar{\cal R}_{\dot\a\dot\b}&=& d\bar{\o}_{\ad\bd}
 +\bar{\o}_{\ad\cd}\wedge\bar{\o}_{\bd}{}^{\cd}
 +e_{\d\ad}\wedge e^{\d}{}_{\bd}\ ,
 \label{radbd}\w2
 {\cal R}_{\a\dot\b}&=&  de_{\a\bd}+ \o_{\a\c}\wedge
 e^{\c}{}_{\bd}+\bar{\o}_{\bd\dd}\wedge e_{\a}{}^{\dd}\
 ,\label{rabd}
 \eea
and
 \be
 {\cal R}^{ab}\ =\ d\o^{ab}+\o^a{}_c\wedge\o^{cb} +\lambda^2
 e^a\wedge e^b\ ,\qquad {\cal R}^a\ =\ d e^a+\o^a{}_b\wedge e^b\ .
 \label{curvcomp} \ee
%


\scs{Further Notation Used for the Solutions}\label{AppDef}


The gauge function $L(x;y,\yb)$ defined in \eq{wL} can be written as
\bea L &=& {2h\over 1+h} \exp (-iy a \yb)\ , \label{L}\eea
where
\bea a_{\a\ad}&=&{\lambda x_{\a\ad}\over 1+h}\ ,\qquad
x_{\a\ad}=(\s^a)_{\a\ad} x_a\ ,\label{aaad}\w2
x^2 &=&  \eta_{ab} x^a x^b\ ,\qquad  h =  \sqrt{1-\lambda^2x^2}\
.\label{aad} \eea
Useful relations that follow from these definitions are
\be a^2={1-h\over 1+h}\ ,\qquad h={1-a^2\over 1+a^2}\ .\ee
The Maurer-Cartan form based on $L$ defined in \eq{wL} yields the
the vierbein and Lorentz connection
\be e_{(0)}{}^{\a\ad}\ =\ -{\l(\s^a)^{\a\ad}dx_a\over h^2}\ ,\qquad
\o_{(0)}{}^{\a\b}\ =\ - {\l^2(\s^{ab})^{\a\b} dx_a x_b\over h^2}\
,\label{adseo}\ee
with Riemann tensor given by
\be
 R_{(0)\m\n,\r\s}\ = \
 -\lambda^2 \left( g_{(0)\mu\rho} g_{(0)\nu\sigma}-
  g_{(0)\nu\rho} g_{(0)\mu\sigma} \right)\ .
\ee
A further useful definition is
\bea b_{\a\b}&=&2\l_{(\a}\m_{\b)}\ ,\qquad \l^\a \mu_\a\ =\ \ft{i}2\
. \label{bab}\eea
It obeys the relation $(b^2)_\a{}^\b\ =\ -\ft 14 \d_\a^\b$ and it
defines an almost complex structure via the relations (see, for
example, \cite{Pope:1982ad})
\be b_{\a\b}=\frac{1}{8} (\sigma^{ab})_{\a\b}\,J_{ab}\ ,\qquad
J_{ab}=(\sigma_{ab})^{\a\b}\,b_{\a\b}\ ,\qquad J_a{}^c J_c{}^b=
-\delta_a^b\ . \label{jj}\ee
Similarly, using the definition
\be {\tilde b}_{\a\b}= a^{-2} (a{\bar b} \bar a)_{\a\b}\
,\label{btilde}\ee
we have the relations
\be {\tilde b}_{\a\b}=\frac{1}{8} (\sigma^{ab})_{\a\b}\,{\tilde
J}_{ab}\ ,\qquad {\tilde J}_{ab}=(\sigma_{ab})^{\a\b}\, {\tilde
b}_{\a\b}\ ,\qquad {\tilde J}_a{}^c {\tilde J}_c{}^b= -\delta_a^b\
.\label{jtilde} \ee
Finally, we have the following definition for spinors used in
describing a Type 3 solution:
\be U_{\ad}={x^a\over \sqrt{x^2}}
\left({\bar\sigma}_a\lambda\right)_{\ad}\ , \qquad V_{\ad}=
{x^a\over \sqrt{x^2}} \left({\bar\sigma}_a \mu\right)_{\ad}\ .
\label{UV}\ee


\scs{Weyl-ordered Projectors}\label{AppProj}


Weyl-ordered projectors $P(y,\bar y)$ can be constructed by
recombining $(y,\bar y)$ into a pair of Heisenberg oscillators
$(a_i,b^j)$ ($i,j=1,2)$ obeying
\bea [a_i,b^j]_\star&=& \d_i^j\ .\eea
For example, one can take
\bea a_1&=&u\ =\ \l^\a y_\a\ ,\qquad b^1\ =\ v\ =\ \mu^\a y_\a\
,\\[5pt]
a_2&=&\bar u\ =\ \bar\l^{\ad}\bar y_{\ad}\ ,\qquad b^2\ =\ \bar v\
=\ \bar\mu^{\ad}\bar y_{\ad}\ ,\eea
where the constant spinors are normalized as
\bea \l^\a\mu_\a&=&\ft{i}2\ ,\qquad \bar\l^{\ad}\bar\mu_{\ad}\ =\
\ft{i}2\ .\eea
The projectors, obeying the appropriate reality conditions, take the
form
\bea P&=&\sum_{n_1,n_2\in~\integ+\ft12}\th_{n_1,n_2} P_{n_1,n_2}\
,\qquad \bar P\ =\ \sum_{n_1,n_2\in~\integ+\ft12}\bar\th_{n_1,n_2}
P_{n_1,n_2}\ ,\label{generalP}\eea
where $\theta_{n_1,n_2}\in\{0,1\}$ and
$\bar\th_{n_1,n_2}\in\{0,1\}$, with
\bea \mbox{$(3,1)$ signature}&:& \th_{n_1,n_2}\ =\
\bar\th_{n_1,n_2}\ ,\\[5pt]
\mbox{$(4,0)$ and ($2,2)$ signatures}&:& \th_{n_1,n_2}\
,\bar\th_{n_1,n_2}\quad\mbox{independent}\ ,\eea
and
\bea P_{n_1,n_2}&=& 4(-1)^{n_1+n_2-\ft{\e_1+\e_2}2}e^{-2\sum_i \e_i
w_i}L_{n_1-\ft{\e_1}2}(4\e_1 w_1)L_{n_2-\ft{\e_2}2}(4\e_2 w_2)\ ,\\[5pt]
w_i&=&b^i a_i\ =\ b^i\star a_i+\ft12\ =\ a_i\star b^i-\ft12\qquad
\mbox{(no sum)}\ ,\eea
with $\e_i=n_i/|n_i|$ and $L_n(x)={1\over n!}e^x{d^n\over
dx^n}(e^{-x}x^n)$ are the Laguerre polynomials. The projector
property follows from
\bea P_{m_1,m_2}\star P_{n_1,n_2}&=& \d_{m_1 n_1}\d_{m_2
n_2}P_{n_1,n_2}\
,\\[5pt]
(w_i-n_i)\star P_{n_1,n_2}&=& 0\ ,\\[5pt] \t(P_{n_1,n_2})&=& P_{-n_1,-n_2}\ .\label{tauP}\eea
Here, $w_i-\ft12$ is the Weyl-ordered form of the number operator,
and
\bea 2(-1)^{n_i-\ft{\e_i}2} e^{-2\e_i w_i}L_{n_i-\ft{\e_i}2}(4\e_i
w_i)&=&\mx{\{}{ll}{|n_i\rangle\langle n_i|&\mbox{for $n_i>0$}\\[5pt]
(-1)^{-n_i-\ft12}|n_i\rangle\langle n_i|& \mbox{for $n_i<0$}}{.}\eea
where
$|n_i\rangle=\ft{(b^i)^{n_i-\ft12}}{\sqrt{(n_i-\ft12)!}}|0\rangle$
with $n_i>0$ belongs to the standard Fock space, built by acting
with $b^i$ on the ground state $|0\rangle$ obeying $a_i|0\rangle=0$,
while
$|n_i\rangle=\ft{(a_i)^{-n_i-\ft12}}{\sqrt{(-n_i-\ft12)!}}|\tilde
0\rangle$ for $n_i<0$ are anti-Fock space states, built by acting
with $a_i$ on the anti-ground state $|\tilde 0\rangle=0$ obeying
$b^i|\tilde 0\rangle=0$. Formally, the inner product between a Fock
space state and an anti-Fock space state vanishes. However, the
corresponding Weyl-ordered projectors have divergent
$\star$-products, as can be seen from the lemma
\bea e^{suv}\star e^{tuv}&=& {1\over
1+\frac{st}4}\exp\left({\frac{s+t}{1+\frac{st}4}\,uv}\right)\ .\eea
Thus, lacking, at present, a suitable regularization scheme that
does not violate associativity and other basic properties of the
$\star$-product algebra, we shall restrict our attention to
projectors that are constructed in either the Fock space or the
anti-Fock space, \emph{i.e.}
\bea \th_{n_1,n_2}&=&1 \quad \mbox{only if $(n_1,n_2)\in Q$}\ ,\eea
where $Q$ is anyone of the four quadrants in the $(n_1,n_2)$ plane.
From \eq{tauP}, it follows that these projectors are not invariant
under the $\tau$ map, and therefore the master fields Type 2 and
Type 3 solutions will be those of the non-minimal model, where the
$\tau$ conditions are relaxed to $\pi\bar\pi$ conditions, which are
certainly satisfied.

We also note that in order to solve the higher-spin equations it is
essential that
\bea [P,\bar P]_\star&=&\sum_{n_1,n_2}
(\th_{n_1,n_2}\bar\th_{n_1,n_2}-\bar\th_{n_1,n_2}\th_{n_1,n_2})P_{n_1,n_2}\
=\ 0\ ,\eea
which holds for independent $\th_{n_1,n_2}$ and
$\bar{\th}_{n_1,n_2}$ parameters (in the Euclidean and Kleinian
signatures). Moreover, one can work with a reduced set of
oscillators, say $a_1=u$ and $b^1=v$, by summing over all values of
$n_2$ using
\bea \sum_{k=0}^\infty t^k L_k(x)&=&(1-t)^{-1}\exp(-xt(1-t)^{-1})\
.\label{generating}\eea
Setting $n=n_1$ and $\e=\e_1$, this leads to
\bea P&=& \sum_{n\in ~\integ+\ft12}\th_n P_n\ ,\qquad \bar P\ =\
 \sum_{n\in ~\integ+\ft12}\bar\th_n P_n\\[5pt]
P_n&=& 2(-1)^{n-\ft{\e}2} e^{-2\e uv} L_n(4\e uv)\
,\label{reduced}\eea
with suitable reality conditions on the $\th_n$ parameters. Finally,
using \eq{generating} once more, one finds that setting all
$\th$-parameters equal to $1$ gives $P=1$.


\scs{Calculation of $V=L^{-1}\star P\star L$}\label{App2}


In this Appendix we compute $V=L^{-1}\star P\star L$ where $L$ is
the gauge function given in \eq{L} and $P$ is a projector of the
form given in \eq{generalP}. Let us begin by considering the case of
$P=P_{\ft12}=2e^{-2uv}$, \emph{i.e.}
\bea V &=&{8h^2\over(1+h)^2} e^{iy a \bar y}\star e^{yby}\star
e^{-iy a\bar  y}\ ,\eea
where $y a\bar y= y^\a a_{\a}{}^{\ad}\bar y_{\ad}$ and $yby=y^\a
b_\a{}^\b y_\b$, with $a_{\a\ad}$ and $b_{\a\b}$ given by \eq{aaad}
and \eq{bab}. The first $\star$-product can be performed treating
the integration variables $(\xi_\a,\eta_\a)$ and
$(\bar\xi_{\ad},\bar\eta_{\ad})$ as separate real variables. Using
the formulae (B.1) provided in \cite{Sezgin:2005pv}, we find
\bea V &=&{8h^2\over(1+h)^2} e^{i ya\bar y+ (y-\bar y a)b(y+a\bar
y)}\star e^{-i y a\bar y}\ .\eea
The remaining $\star$-product leads to the Gaussian integral
\bea V&=&{8h^2\over(1+h)^2} \int {d^4\xi d^4\eta\over (2\pi)^4}
e^{\ft12 \Xi^I M_I{}^J \Xi_J+\Xi^I N_I +(y-\yb a)b(y+a\yb)}\
,\label{b4} \eea
where $\Xi^I=(\x^\a,\bar \x^{\ad};\eta^{\a},\bar\eta^{\ad})$ and
$\Xi_I=(\x_\a,\bar
\x_{\ad};\eta_{\a},\bar\eta_{\ad})=\Xi^J\Omega_{JI}$, with
block-diagonal symplectic metric $\Omega=\e\oplus\bar
\e\oplus\e\oplus\bar \e$, and
\bea M&=&\left[\ba{cc}A&-i\\i&B\ea\right]\ ,\\[5pt] A&=&
\left[\ba{cc}2b&ia+2ba\\ia-2ba&-2\bar aba\ea \right]
\ ,\qquad B\ =\ \left[\ba{cc}0&-ia\\-i\bar a&0\ea\right]\ ,\\[5pt] N&=&
\left[\ba{l} i(1-2ib)a\bar y +2by\\-2\bar aba\bar y +i\bar
a(1+2ib)y\\-ia\bar y\\-i\bar ay\ea \right]\ .\eea
The Gaussian integration gives
\bea V &=&{8h^2\over(1+h)^2\sqrt{\det M}}~e^{\ft12 N^I(M^{-1})_I{}^J
N_J  + (y-\bar y a)b(y+a\bar y)}\ .\eea
From $\det M= \det (1+AB)$, and noting that the matrices defined as
\bea C&\equiv & {BA-a^2\over 2i}\ =\ \left[\ba{cc}a^2b&a^2ba\\-\bar
ab&-\bar aba\ea\right]\ ,\qquad \tilde C\ \equiv \ {AB-a^2\over 2i}\
=\ \left[\ba{cc}-a^2b&-ba\\\bar aba^2&\bar aba\ea\right]\ ,\eea
are nilpotent, i.e. $C^2 = \tilde C^2\ =\ 0$, one finds
\bea\det M&=& (1-a^2)^4\ ,\eea
and, using $1-a^2=2h/(1+h)$, the pre-factor in $V$ is thus given by
\bea {8h^2\over(1+h)^2\sqrt{\det M}}&=&2\ .\eea
Next, using geometric series expansions, one finds
\bea M^{-1}&=& {i\over (1-a^2)}\left[\ba{cc}i(1-a^2)B+2B\tilde C&
-(1-a^2)-2iC\\ 1-a^2+2i\tilde C&i(1-a^2)A+2AC\ea\right]\ ,\eea
and
\bea \ft12 N^I(M^{-1})_I{}^J N_J&=& {4a^2 yby +2(1+4a^2-a^4) yba\yb
-(3-a^2)(1+a^2) \bar y \bar aba\bar y\over (1-a^2)^2}\ .\eea
Adding the classical term in the exponent in \eq{b4} yields the
final result
\bea V &=& 2 \exp \left(-{ [2\yb \bar a -(1+a^2)y]\,b\,[2a\yb
+(1+a^2)y]\over (1-a^2)^2}\right)\ .\label{Vrho0}\eea
The projector property $V\star V=V$ follows manifestly from
\bea V&=& 2\exp (-2\tilde u\tilde v)\ ,\qquad [\tilde u,\tilde
v]_\star\ =\ 1\ ,\eea
where
\bea \tilde u&=&\l^\a\eta_\a\ ,\qquad \tilde v\ =\ \mu^\a\eta_\a\
,\eea
with
\bea \eta_\a&=& {[(1+a^2)y+2a\yb]_\a\over 1-a^2}\ ,\qquad
[\eta_\a,\eta_\b]_\star\ =\ 2i\e_{\a\b}\ .\eea
Thus, the net effect of rotating the projector $P_{\ft12}(u,v)$
given in \eq{reduced} is to replace the oscillators $u$ and $v$ by
their rotated dittos $\tilde u$ and $\tilde v$. We claim, without
proof, that this generalizes to any $n$, \emph{viz.}
\bea L^{-1}\star P_n(u,v)\star L&=& P_n(\tilde u,\tilde v)\ .\eea
Similarly, for $P_n(\bar u,\bar v)$ we have
\bea L^{-1}\star P_n(\bar u,\bar v)\star L&=& P_n(\tilde{\bar
u},\tilde {\bar v})\ ,\eea
where
\bea \tilde{ \bar u}&=&\bar\l^{\ad}\bar \eta_{\ad}\ ,\qquad \tilde{
\bar v}\ =\ \bar\mu^{\ad}\bar\eta_{\ad}\ ,\eea
with
\bea \bar\eta_{\ad}&=& {[(1+a^2)\yb+2\bar a y]_{\ad}\over 1-a^2}\
,\qquad [\bar\eta_{\ad},\bar\eta_{\bd}]_\star\ =\ 2i\e_{\ad\bd}\
.\eea
Finally, using $[\eta_{\a},\bar\eta_{\ad}]_\star=0$, we deduce that
\bea V&=&L^{-1}\star P\star L\ =\ \sum_{n_1,n_2}\th_{n_1,n_2}
P_{n_1,n_2}(\tilde u,\tilde v;\tilde{\bar u},\tilde{\bar v})\
.\label{generalV}\eea
%


\end{appendix}

\newpage


\end{document}